\documentclass[twocolumn,english,prb,floatfix,superscriptaddress,twocolumn,showpacs,amsmath,amssymb,reprint]{revtex4}
\usepackage[T1]{fontenc}
\usepackage[latin9]{inputenc}
\usepackage{color}
\definecolor{shadecolor}{rgb}{1, 0, 0}
\usepackage{verbatim}
\usepackage{float}
\usepackage{framed}
\usepackage{amsmath}
\usepackage{amssymb}
\usepackage{graphicx}

\makeatletter
\PassOptionsToPackage{caption=false}{subfig}
\usepackage{hyperref}
\hypersetup{
breaklinks=true,
colorlinks=true,
citecolor=blue,
linkcolor=blue,
filecolor=blue,
urlcolor=blue
}
\IfFileExists{lmodern.sty}{\usepackage{lmodern}}{}
\makeatother

\usepackage{babel}

\begin{document}

\title{Influence of Landau levels in the phonon dispersion of Weyl semimetals}

\author{P. Rinkel}

\affiliation{D\'epartement de Physique, Institut Quantique and Regroupement Qu\'eb\'ecois
sur les Mat\'eriaux de Pointe, Universit\'e de Sherbrooke, Sherbrooke,
Qu\'ebec, Canada J1K 2R1}

\author{P. L. S. Lopes}

\affiliation{D\'epartement de Physique, Institut Quantique and Regroupement Qu\'eb\'ecois
sur les Mat\'eriaux de Pointe, Universit\'e de Sherbrooke, Sherbrooke,
Qu\'ebec, Canada J1K 2R1}
\affiliation{Department of Physics and Stewart Blusson Institute for Quantum Matter, University of British Columbia, Vancouver, Canada V6T 1Z1}

\author{Ion Garate}

\affiliation{D\'epartement de Physique, Institut Quantique and Regroupement Qu\'eb\'ecois
sur les Mat\'eriaux de Pointe, Universit\'e de Sherbrooke, Sherbrooke,
Qu\'ebec, Canada J1K 2R1}

\date{\today}
\begin{abstract}
Weyl semimetals display unusual electronic transport properties when placed under magnetic fields. 
Here, we investigate how magnetic fields alter the dynamics of long wavelength lattice vibrations in these materials.
To that end, we develop a theory for the phonon dispersion, which incorporates contributions from chiral and nonchiral Landau levels, electron-phonon interactions,  electron-electron interactions, and disorder.
We predict (i) a magnetic-field-induced hybridization between optical phonons and plasmons, 
(ii) avoided crossings between pseudoscalar optical phonons and electronic excitations originating from nonchiral Landau levels,
(iii) a sharp dependence of the sound velocity on the relative angle between the sound propagation and the magnetic field.
%(iv) quantum oscillations in the sound velocity, with an amplitude that has an unusual dependence on the magnetic field.
We compare our results to recent theoretical studies on the signatures of the chiral anomaly in phonon dynamics.
%{\bf in Mermin formula we neglect internode scattering, I think}
\end{abstract}
\maketitle

\section{Introduction}

%{\bf the sign of $\Pi_A$ depends on the sign of ${\bf q}\cdot{\bf B}$}
%{\bf distinguish different meanings of Born charge not to confuse the reader}

Weyl semimetals (WSM) are three dimensional crystals whose low-energy excitations consist of topologically protected  Weyl fermions.\cite{Armitage}
In recent years, these materials have attracted a considerable interest due to their unusual electronic properties that are connected to concepts of particle physics and topology.
The existence of Fermi arcs at the surface of the sample is one example of what makes WSM special.
Another characteristic effect, of particular interest in this paper, is the chiral anomaly.\cite{Kharzeev} 
Numerous theoretical and experimental papers (see e.g. Refs. [\onlinecite{Huang, Hirschberger, Goswami, Arnold, Reis, Zhou, Zhang, Panf, Hutasoit, Redell, Grushin, Potter, Souma}]) have been devoted to understand and confirm these peculiar electronic phenomena in WSM.

More recently, there have been attempts at finding interesting {\em nonelectronic} phenomena in WSM.
Less developed than the study of electronic properties, this line of research may nevertheless be useful in order to understand the wider implications of Weyl fermions in physical observables.
Thus, various authors have studied the phonon Hall viscosity,\cite{Shapourian, Cortijo} sound attenuation,\cite{Andreev, Pikulin} and the dispersion of optical phonons\cite{Song, Us} in WSM. 

The present work is motivated mainly by the results of Refs. [\onlinecite{Song,Us}].
These papers predicted a new signature of the chiral anomaly in lattice dynamics, in the form of a magnetic-field-induced effective charge for pseudoscalar phonons.
This signature does not suffer from current jetting effects that add difficulty to the interpretation of electronic transport measurements of the chiral anomaly.
By definition, pseudoscalar phonons create deformation potentials of equal magnitude and opposite sign for Weyl nodes of opposite chirality.
Also, the effective charge of a phonon mode is defined as the electric dipole moment per unit cell caused by the lattice vibration.
According to Refs. [\onlinecite{Song,Us}], a pseudoscalar optical phonon that is infrared inactive (i.e. without an effective charge) at a vanishing magnetic field becomes infrared active in a magnetic field. 
For the same reason, this pseudoscalar phonon can hybridize with electronic collective excitations, giving raise to a polariton-like avoided crossing in its dispersion.  
These effects can be attributed to the chiral anomaly.

The analysis in Ref. [\onlinecite{Song}] was largely phenomenological, without a microscopic calculation of response functions.
In Ref. [\onlinecite{Us}], a microscopic theory was developed perturbatively (to first order) in the external magnetic field, and at negligible chemical potential, temperature and disorder.
However, because of the linear response approximation, the hybridization gap for the optical phonon could not be reliably estimated at strong magnetic fields.
Likewise, within the linear response approximation, the effect of the magnetic field in the electronic screening was ignored.
This gave rise to a non-plasmonic pseudoscalar boson that hybridized with the optical phonon.
But such result is no longer reliable at strong fields, which strongly modify the dielectric function. 
In view of these limitations, it is desirable to extend Refs. [\onlinecite{Song,Us}] by developing a microscopic theory for the phonon dispersion in WSM, which will be non perturbative in the magnetic field and will include the effects of finite chemical potential, temperature and disorder. 
That is the objective of the present paper.

We begin in Sec. \ref{sec:phodis} with a general formalism that describes the impact of Weyl fermions in the phonon dispersion.
The formalism, applicable to an arbitrary strength and direction of the magnetic field, is based on a functional integral approach.
In this approach, Weyl fermions are integrated out in order to obtain an effective action for the phonons.
The procedure being standard but tedious, the details of the calculation are relegated to the Appendices.
For simplicity, we limit our discussion to pseudoscalar and scalar phonons (the latter generate deformation potentials that are equal in magnitude and sign for Weyl nodes of opposite chirality).

In Sec. \ref{sec:optical}, we apply the formalism from Sec. \ref{sec:phodis}  to analyze the dispersion of optical phonons. 
The main prediction from this section is an avoided crossing between the plasmon and the optical phonon.
The hybridization can be controlled by the strength and direction of the magnetic field.
Though this result is in qualitative agreement with the results from Refs. [\onlinecite{Song,Us}], the novelty here is that the avoided crossing can also take place for scalar phonons.
We estimate the hybridization gap and find it to be small but potentially observable in high precision experiments and at strong magnetic fields.
In the quantum limit, the gap scales as the square root of the magnetic field, which is in apparent disagreement with the result from Ref.~[\onlinecite{Song}].
Similarly, we correct two inaccurate statements made in Ref. [\onlinecite{Us}]: (i) all phonons (rather than just $A_1$ phonons) are able to scatter Weyl fermions in the long-wavelength limit; (ii) the pseudoscalar electron-phonon coupling can be nonzero even in crystals that possess mirror symmetries.

In Sec. \ref{sec:acoustic}, we apply the formalism from Sec. \ref{sec:phodis} to the study of acoustic phonons in a piezoelectric WSM. 
We show that the contribution from chiral Landau levels to the group velocity of acoustic waves is sharply dependent on the relative angle between the magnetic field and the direction of sound propagation.
Also, an increase in the magnetic field strength results in a decrease of the sound velocity.
This effect has the same origin as the well-known negative magnetoresistance,\cite{Huang,Andreev} which is in turn related to the chiral anomaly. 
At moderate magnetic fields, our theory predicts quantum oscillations in the sound velocity.
%, with an amplitude growing as $B^{3/2}$ with the field.   

Section \ref{sec:summ} summarizes the main results and the subsequent Appendices collect the technical details supporting the main results.

\section{Phonon dispersion in a magnetic field}
\label{sec:phodis}

The objective of this section is to set the formalism for the phonon dispersion in a WSM placed under a magnetic field. 
We begin with a low-energy effective Hamiltonian of two Weyl nodes in an external magnetic field,
\begin{equation}
{\cal H}_{\rm el}^{(0)}=\int d^3r\,\hat{\Psi}^{\dagger}(\mathbf{r})\left[b_0\tau_z+v_F\tau_{z}\boldsymbol{\sigma}\cdot\left(\mathbf{p}-e\mathbf{A}\right)\right]\hat{\Psi}(\mathbf{r}),
\label{eq:Hel}
\end{equation}
where $\mathbf{p}$ is the momentum operator, $\mathbf{A}$ is the electromagnetic vector potential, $v_F$ is the Fermi velocity, and $b_0$ is half of the energy difference between the two Weyl nodes.
In addition, ${\boldsymbol\tau}$ and ${\boldsymbol\sigma}$ are vectors of Pauli matrices, with $\tau_z$ labelling the opposite chiralities of the two nodes 
and $\sigma_z$ denoting the two degenerate states at each Weyl node.
Also, $\hat{\Psi}({\bf r})$ is a four-component spinor  whose elements 
$\hat{\Psi}_{\sigma\tau}({\bf r})$ are operators that annihilate a Weyl fermion at position ${\bf r}$ and node $\tau$  in a state $\sigma$ ($\sigma=\pm$ and $\tau=\pm$ are the eigenvalues of $\sigma_z$ and $\tau_z$, respectively).

In the presence of a static and uniform external magnetic field $B\hat{\bf z}$, the electronic structure corresponding to Eq.~(\ref{eq:Hel}) consists of
Landau levels (LLs) that disperse along the direction of the magnetic field. 
The energy bands are \cite{Burkov}
\begin{align}
E_{n k_{z} \tau} &= b_0\tau+{\rm sgn}(n)\tau\hbar v_F\sqrt{k_{z}^{2}+2|n|/\ell_{B}^{2}} \, \, (\text{for } n\neq 0)\nonumber\\
E_{0 k_{z} \tau}&=b_0\tau-\tau\hbar v_F k_{z}\,\, (\text{for } n=0),
\end{align}
where $n\in\mathbb{Z}$ is the LL index and $l_B  = [\hbar/ (e B)]^{1/2}$ is the magnetic length.
The corresponding eigenstates will be denoted as $| n k_z  \tau\rangle$ (for brevity we omit the labels for the centers of the cyclotron orbits, which will be incorporated below through a macroscopic degeneracy factor).
The two $n=0$ LLs are chiral, in the sense that their group velocities are determined by the chirality and have a fixed sign regardless of $k_z$.
In contrast, the $n\neq 0$ LLs are nonchiral (their group velocities change sign when $k_z$ changes sign). 
The chiral LLs are responsible for the chiral anomaly in WSM; one of the main objectives below will be to analyze their impact on the phonon dispersion.

Phonon properties are sensitive to magnetic fields mostly due to electron-phonon interactions.
In this work, we will concentrate on long wavelength phonons, whose wave vectors are smaller than the separation between Weyl nodes in momentum space at zero magnetic field.\cite{wavelength}
The coupling of these phonons to Weyl fermions is therefore diagonal in the node index $\tau$, though the magnitude of the coupling can be different for $\tau=+$ and $\tau=-$. 
If the momentum at which a Weyl node is located contains no symmetry (i.e., if the little group therein contains only the identity), then all phonon modes are able to induce electron scattering in its vicinity.
This statement, justified in Appendix A, corrects an erroneous claim made in Ref. [\onlinecite{Us}].
Some of the long-wavelength phonons can in addition scatter a Weyl fermion from one value of $\sigma$ to another.
However, for the sake of brevity we will hereafter focus on electron-phonon couplings that are diagonal in $\sigma$ and have the same magnitude for $\sigma=\pm$.
%; these couplings will suffice to obtain the main physical effect that results of this work.
In sum, the electron-phonon interaction we work with can be written as 
\begin{equation}
{\cal H}_{\rm el-ph}=\int d^3 r\,\hat{\Psi}^{\dagger}(\mathbf{r})\left[\phi_{0}(\mathbf{r},t)+\tau_{z}\phi_{z}(\mathbf{r},t)\right]\hat{\Psi}(\mathbf{r}),\label{eq:Hep}
\end{equation}
where 
\begin{comment}
\begin{align*}
\phi_{0}(\mathbf{r)} & =\frac{1}{2}\sum_{\sigma\tau\mathbf{q}}\text{e}^{i\mathbf{q}\cdot\mathbf{r}}v_{\mathbf{q}}\frac{1}{\sqrt{N}\mathcal{V}_{cell}}\sum_{s}\int_{Cell}\text{d}^{3}\mathbf{r}^{'}\,\left|u_{\sigma\tau}(\mathbf{r})\right|^{2}\mathbf{p}_{\mathbf{q}\lambda s}\cdot\frac{\partial U(\mathbf{r}^{'}-\mathbf{t}_{s})}{\partial\mathbf{t}_{s}}\text{e}^{-i\mathbf{q\cdot r}^{'}}\\
\phi_{z}(\mathbf{r)} & =\frac{1}{2}\sum_{\sigma\tau\mathbf{q}}\tau\,\text{e}^{i\mathbf{q}\cdot\mathbf{r}}v_{\mathbf{q}}\frac{1}{\sqrt{N}\mathcal{V}_{cell}}\sum_{s}\int_{Cell}\text{d}^{3}\mathbf{r}^{'}\,\left|u_{\sigma\tau}(\mathbf{r})\right|^{2}\mathbf{p}_{\mathbf{q}\lambda s}\cdot\frac{\partial U(\mathbf{r}^{'}-\mathbf{t}_{s})}{\partial\mathbf{t}_{s}}\text{e}^{-i\mathbf{q\cdot r}^{'}}
\end{align*}
\end{comment}
$\phi_{0}({\bf r},t)$ and $\phi_{z}({\bf r},t)$ are the scalar and pseudoscalar deformation potentials (see Ref.~[\onlinecite{Us}] and Appendix A). 
The scalar deformation potential induces a global shift in energy for the two Weyl nodes, while the pseudoscalar deformation potential induces a relative shift in energy between them. 
In other words, a scalar (pseudoscalar) phonon couples with the same magnitude and same (opposite) sign to Weyl nodes of opposite chirality.

Such deformation potentials may be related to the normal mode coordinate $v_{{\bf q}\lambda}(t)$ of the lattice vibrations via
\begin{equation}
\phi_{0(z)}(\mathbf{r},t)=\sum_{\mathbf{q}\lambda} e^{i\mathbf{q}\cdot\mathbf{r}}g^\lambda_{0(z)}(\mathbf{q})v_{\mathbf{q}\lambda}(t),
\end{equation}
where ${\bf q}$ is the phonon momentum  and $g^\lambda_{0(z)}$ is the electron-phonon coupling for a scalar (pseudoscalar) phonon mode labeled as $\lambda$.
The sum over ${\bf q}$ has an ultraviolet cutoff, so that $\phi_{0(z)}$ varies slowly with ${\bf r}$.
In our convention, $g$ scales as $1/\sqrt{N}$, where $N$ is the number of unit cells in the crystal (see Ref. [\onlinecite{Us}] and Appendix A).
From the reality of lattice displacements, $g_{0(z)}^\lambda({\bf q}) = [g_{0(z)}^\lambda(-{\bf q})]^*$ and $v_{{\bf q}\lambda}=v_{-{\bf q}\lambda}^*$. 
For optical phonons, $g_z$ and $g_0$ remain finite in the $q\to 0$ limit, while they vanish for acoustic phonons.
For simplicity, from now on we concentrate on a single phonon mode and drop the label $\lambda$ from $v$ and $g$.

From a symmetry point of view, long-wavelength scalar phonons are present in all crystals; they are the so-called $A_1$ phonons, whose polarization vectors remain invariant under all operations of the crystal's point group.
Pseudoscalar phonons likewise exist in all crystal point groups, though they do not necessarily {\em occur} in all crystals.\cite{Song}
For example, in crystals belonging to the $C_{4v}$ point group, $A_2$ phonons are pseudoscalar; yet, TaAs (whose point group is $C_{4v}$) does not have $A_2$ phonons.\cite{Liu} 
A complete list of pseudoscalar phonons in different crystalline point groups can be found in Ref.~[\onlinecite{Anastassakis}].
In addition, it has very recently been predicted that magnetic fields might induce significant pseudoscalar electron-phonon interactions in crystals that would otherwise lack pseudoscalar phonons.\cite{Kim}

In crystals containing mirror planes, ${\bf q}\simeq 0$ pseudoscalar phonons are invariant under all point group operations except those involving mirror reflections, under which they are odd (hence the name of pseudoscalar). 
As we show in Appendix A (see also Ref.~[\onlinecite{Song}] for an alternative proof), these modes couple to electrons via $g_0=0$ and $g_z\neq 0$.
Constrastingly, ${\bf q}\simeq 0$ scalar phonons are invariant under mirrors and couple to electrons with $g_0\neq 0$ and $g_z=0$.
These statements correct a mistake made in Ref.~[\onlinecite{Us}], which erred by claiming that $g_z\neq 0$ is possible only in crystals lacking mirror planes (i.e., in enantiomorphic crystals). 
As it turns out, in enantiomorphic crystals there is no distinction between scalar and pseudoscalar phonons; $A_1$ phonons couple to electrons with  $g_0\neq 0$ and $g_z\neq 0$ in this case.
In Eq.~(\ref{eq:Hel}), the breaking of mirror symmetries is modeled by $b_0\neq 0$.
Most of the numerical results shown below will be obtained for $b_0=0$, which implies that $g_z=0$ ($g_0=0$) for scalar (pseudoscalar) phonons. 

Besides the electron-phonon interaction, another important ingredient for the phonon dispersion is the Coulomb interaction between Weyl fermions.
This interaction, responsible for the screening of the electron-phonon interaction, can be modeled from the Hamiltonian 
\begin{equation}
\label{eq:Hee}
{\cal H}_{\rm el-el} = \int d^3 r\, d^3 r'\, V({\bf r}-{\bf r}') : \rho({\bf r}) \rho({\bf r}') :,
\end{equation}
where  $\rho({\bf r}) = \hat{\Psi}^\dagger({\bf r}) \Psi({\bf r})$ is the electronic density operator, the colon stands for normal ordering,  $V(r)=e^2/(4\pi\varepsilon_\infty r)$, and $\varepsilon_\infty$ is a phenomenological parameter capturing the screening due to high-energy (non-Weyl) fermions. 

Combining Eqs.~(\ref{eq:Hel}), (\ref{eq:Hep}) and (\ref{eq:Hee}), one can calculate the phonon dispersion in a WSM.
We carry out this calculation using a functional integral formalism; the procedure is standard and the details are shown in Appendix A.
One of the central quantities in the theory is the electronic Green's function matrix $G_0(x, x')$ obtained from Eq.~(\ref{eq:Hel}), where $x=({\bf r},t)$ and $t$ is the time coordinate.
This object can be factorized \cite{Schwinger} as $G_0(x,x')=\Phi(x,x') \mathcal{G}_0(x-x')$,
where $\mathcal{G}_0$ is the translationally invariant part of $G_0$ and $\Phi$ is a phase factor obeying $\Phi(x,x') \Phi(x',x)=1$.

The outcome of the aforementioned calculation is the following equation for the phonon frequency $\omega$ as a function of the phonon momentum ${\bf q}$:
\begin{widetext}
\begin{equation}
\omega^2= \omega_{{\rm ph},\mathbf{q}}^{2}+\frac{{\cal V}}{M\varepsilon(q)}\Biggl\{\left(\left|g_{0}(\mathbf{q})\right|^{2}+\left|g_{z}(\mathbf{q})\right|^{2}\right)\Pi^R_V(q)+2\text{Re}\left[g_{0}(\mathbf{q})g_{z}(-\mathbf{q})\right]\Pi^R_A(q)-V({\bf q})\left[\Pi^R_V(q)^{2}-\Pi^R_A(q)^{2}\right]\left|g_{z}(\mathbf{q})\right|^{2}\Biggr\},
\label{eq:GenDisp}
\end{equation}
\end{widetext}
where $q=\left(\mathbf{q},\omega\right)$, ${\cal V}$ is the sample volume, $\omega_{{\rm ph},\mathbf{q}}$ is the bare phonon frequency, 
$\Pi^{R}_V$ and $\Pi^R_A$ are retarded density response functions (more on these below), 
$\varepsilon(q) =1-V({\bf q})\Pi^R_V(q)$ is the dielectric function, and $V({\bf q})=e^{2}/(\varepsilon_{\infty}\mathbf{q}^{2})$.
%\end{widetext}
Equation (\ref{eq:GenDisp}) is a central result of this paper.
To be precise, Eq.~(\ref{eq:GenDisp}) applies only to non-polar phonons whose effective charge vanishes in the absence of electron-phonon interactions.
In Appendix A, we derive a generalized version of Eq.~(\ref{eq:GenDisp}) for polar optical phonons. 
For $g_z=0$, Eq.~(\ref{eq:GenDisp})  reduces to the usual expression found in textbooks;\cite{Mahan} the terms involving $g_z$ will nevertheless play an important role in Sec.~\ref{sec:optical}.
%For instance, the last term in the right hand side of Eq. (\ref{eq:GenDisp}) is particularly noteworthy, as it remains finite in the $|{\bf q}|\to 0$ limit.
%This reflects the fact that scalar and pseudoscalar electron-phonon interactions are screened differently by Coulomb interactions.
%in the same way by Coulomb interactions.

The solutions $\omega({\bf q})$ of Eq.~(\ref{eq:GenDisp}) are complex.
The real part of a solution describes the phonon frequency as a function of the phonon momentum, while the imaginary part captures the phonon linewidth.
The origin of the phonon linewidth resides partly in the electron-phonon interaction, through the imaginary part of $\Pi^R_{V(A)}$.
Other sources for the phonon linewidth are the lattice anharmonicities or simply the limited precision of the measurement instrument. 
These extra sources will be taken into account with a phenomenological parameter $\eta$, by replacing $\omega\to \omega+i\eta$ in the left hand side of Eq.~(\ref{eq:GenDisp}). 
Assuming that $\eta$ is small, its contribution to the real part of the phonon frequency will be neglected. 

The functions $\Pi^R_V$ and $\Pi^R_A$ describe the response of the electron density to a scalar and pseudoscalar potential, respectively. 
They can be obtained from their imaginary-time counterparts, 
\begin{align}
\Pi_V(\mathbf{q},i\omega) & =\frac{T}{\mathcal{V}}\sum_{{\bf k},i\nu_n}\text{tr}\left[\mathcal{G}_0(\mathbf{k},i\nu_n)\mathcal{G}_0(\mathbf{k}+\mathbf{q},i\nu_n+i\omega)\right]\nonumber\\
\Pi_A(\mathbf{q},i\omega) & =\frac{T}{\mathcal{V}}\sum_{\mathbf{k},i\nu_n}\text{tr}\left[\tau_z\mathcal{G}_0(\mathbf{k},i\nu_n)\mathcal{G}_0(\mathbf{k}+\mathbf{q},i\nu_n+i\omega)\right]
\label{eq:PIA1},
\end{align}
via analytical continuation ($i\omega \to \omega+i 0^+$).
Here, $T$ is the temperature, $\nu$ is a fermionic Matsubara frequency and $\mathcal{G}_0({\bf k},i\nu)$ is the Fourier transform of the translationally invariant part of the Green's function $\mathcal{G}_0(x)$.
A trace is taken over the $\sigma$ and $\tau$ degrees of freedom.
As shown in Appendix B, Eq.~(\ref{eq:PIA1}) can be recasted into the form 
\begin{widetext}
\begin{align}
\label{eq:pill}
\Pi_V({\bf q},i\omega) &= \frac{1}{4\pi^2 l_B^2} \sum_{n n'}\int \text{d}k_z\sum_{\tau} |\langle n k_z \tau | e^{i {\bf q}\cdot{\bf r}}|n' k_z+q_z \tau\rangle|^2 \frac{f(E_{n k_z \tau}) - f(E_{n' k_z+q_z \tau})}{E_{n k_z  \tau} - E_{n' k_z+q_z  \tau} + i\omega} \nonumber\\
\Pi_A({\bf q},i\omega) &= \frac{1}{4\pi^2 l_B^2} \sum_{n n'}\int \text{d}k_z\sum_{\tau} \tau |\langle n k_z \tau | e^{i {\bf q}\cdot{\bf r}}|n' k_z+q_z \tau\rangle|^2 \frac{f(E_{n k_z  \tau}) - f(E_{n' k_z+q_z  \tau})}{E_{n k_z  \tau} - E_{n' k_z+q_z  \tau} + i\omega}, 
\end{align}
%\end{widetext}
where $f(x)$ is the Fermi-Dirac distribution with a chemical potential $\mu$ and ${\bf q}=({\bf q}_\perp, q_z)$.
The sum over the centers of cyclotron orbits is responsible for the prefactor $1/l_B^2$.
Also,
\begin{equation}
  \langle n k_z \tau | e^{i {\bf q}\cdot{\bf r}}|n' k_z+q_z \tau\rangle = \sqrt{\frac{n_{\rm min}!}{n_{\rm max}!}} e^{-{\bf q}_\perp^2 l_B^2/4} \left(\frac{l_B}{\sqrt{2}}(i q_x + q_y)\right)^{n_{\rm max}-n_{\rm min}} L_{n_{\rm min}}^{n_{\rm max}-n_{\rm min}}({\bf q}_\perp^2 l_B^2/2),
\end{equation}
where $L^\alpha_\beta(x)$ is the Laguerre polynomial (see Appendix B), $n_{\rm max} = {\rm max}(|n|,|n'|)$ and $n_{\rm min} = {\rm min}(|n|,|n'|)$.
When ${\bf q}_\perp=0$,  $ \langle n k_z \tau | e^{i {\bf q}\cdot{\bf r}}|n' k_z+q_z \tau\rangle =\delta_{|n|,|n'|}$.
  \end{widetext}
The sum over Landau levels in  Eq.~(\ref{eq:pill}) can be separated into intraband ($n'=n$) and interband $(n'\neq n)$ parts; we use this nomenclature repeatedly in the following sections.
Appendix B manipulates Eq.~(\ref{eq:pill}) further in order to reach expressions that are more convenient for numerical calculations.

When $b_0=0$, the perturbative expansion of $\Pi_A$ in powers of the magnetic field gives the well-known VVA triangle diagram to leading (first) order in the magnetic field.\cite{Adler}
This diagram,  containing two vector (V) vertices and one axial (A) vertex, describes the axial current induced by collinear electric and magnetic fields and is therefore responsible for the chiral anomaly. 
In contrast, at $b_0=0$, the leading terms in the expansion of $\Pi_V$ are a bubble diagram with two vector vertices and a triangle diagram with three vector vertices, none of which are anomalous. 
Because $\Pi_A$ appears in Eq.~(\ref{eq:GenDisp}) only when $g_z\neq 0$, smoking gun signatures of the chiral anomaly at $b_0=0$ occur only for pseudoscalar phonons. 
As we discuss below, the characteristic signature of the chiral anomaly will be a magnetic-field-induced hybridization between a pseudoscalar optical phonon and the plasmon at zero momentum.
When $b_0\neq 0$, it turns out that the perturbative expansion of $\Pi_V$ in powers of the magnetic field contains the same anomalous term as $\Pi_A$ (see Section 3 of Appendix B).
At any rate, when $b_0\neq 0$, the same phonon mode will couple to electrons via $g_0\neq 0$ and $g_z\neq 0$.

From a physical point of view, $\Pi_V$ and $\Pi_A$ are related to the effective phonon charge induced dynamically by Weyl fermions.
The effective charge for a phonon mode $\lambda$ of momentum ${\bf q}$ is proportional to
$\partial {\bf P}/\partial v_{{\bf q}\lambda}$,
where ${\bf P}$ is the electric polarization per unit cell.\cite{Gonze}
In insulating crystals, the phonon modes with a nonzero effective charge absorb infrared light via the ${\bf P}\cdot{\bf E}_{\rm em}$ coupling, where ${\bf E}_{\rm em}$ is the electric field from the electromagnetic wave.
On the other hand, phonon modes with a vanishing effective charge are infrared inactive.
In conducting systems, itinerant charge fluctuations couple to lattice vibrations via the electron-phonon coupling, endowing them with an electric polarization.
As shown in Appendix A, this results in an additional effective phonon charge ${\bf Q}_{\rm el}$ that obeys the relation
\begin{equation}
\label{eq:bc}
{\bf Q}_{\rm el}(q)\cdot{\bf q} = i\frac{e\cal{V}}{\sqrt{N}}\left[\Pi^R_V(q) g_0({\bf q}) + \Pi^R_A(q) g_z({\bf q})\right].
%{\bf Q}_{\rm el}(q)\cdot{\bf q} \propto e \left[\Pi^R_V(q) g_0({\bf q}) + \Pi^R_A(q) g_z({\bf q})\right],
\end{equation}
The proportionality factor ${\cal V}/\sqrt{N}$ appears from particular conventions employed for e.g. Fourier transforms, and ensures that ${\bf Q}_{\rm el}$ is an intensive quantity.
The first and second terms on the right hand side of Eq.~(\ref{eq:bc}) capture the Weyl fermion contribution to the effective charge of scalar and pseudoscalar phonons, respectively.
Although it is often stated that itinerant electrons screen preexisting effective phonon charges, it is less recognized that itinerant electrons can at the same time produce phonon effective charges that would be absent in the insulating phase.
This idea, which is not unique to WSM, will be pertinent in Sec.~\ref{sec:optical}.

Up until now, we have ignored the presence of disorder. 
Equation~(\ref{eq:GenDisp}) remains formally valid in a disordered system (provided one uses disorder-averaged Green's functions and neglects impurity vertex corrections), but Eqs.~(\ref{eq:PIA1}) and (\ref{eq:pill}) do not.
The use of Eqs.~(\ref{eq:PIA1}) and (\ref{eq:pill}) can still be justified for optical phonons, by assuming that the phonon frequency 
exceeds the scattering rate off impurities.
In contrast, Eqs.~(\ref{eq:PIA1}) and (\ref{eq:pill}) need to be modified for long wavelength acoustic phonons, whose frequencies can easily be small compared to the impurity scattering rate.
Below, we will neglect disorder when discussing optical phonons, but not when discussing acoustic phonons.

\section{Optical phonons}
\label{sec:optical}

In this section, we investigate the optical phonon dispersion in the strong and intermediate magnetic field regimes. 
In order to better highlight the contribution from the chiral anomaly, we will often compare results between purely scalar and purely pseudoscalar phonons.
As mentioned above, this distinction is well justified at $b_0=0$.
% this situation will be briefly discussed.

\subsection{Quantum limit\label{subsec:Quantum-limit}}

In the quantum limit,\cite{Miura} the magnetic field $B$ is sufficiently strong that the energy separation between the chiral LLs and their closest nonchiral LLs exceeds $\mu$.
Accordingly, $\mu$ intersects only the chiral LLs.
This situation corresponds to $B>B_{\rm QL}$, where 
\begin{equation}
B_{\text{QL}}\equiv {\rm max}\left\{\frac{(\mu+b_0)^{2}}{2v_F^{2}e\hbar}, \frac{(\mu-b_0)^{2}}{2v_F^{2}e\hbar}\right\}. 
\end{equation}
For $|\mu\pm b_0|\simeq 20 {\rm meV}$ and $v_F = 2\times 10^5 {\rm m/s}$,\cite{Lee} $B_{QL}\simeq 10 {\rm T}$ is a readily attainable field.

Deep enough into the quantum limit, the electronic response functions from Eq.~(\ref{eq:pill}) can be approximated by keeping solely the chiral LL contribution ($n=n'=0$). 
The result reads
\begin{align}
\Pi^R_V\left(q\right) & \simeq\frac{\ell_{B}^{-2}}{2\pi^{2}\hbar v_F}\,\frac{v_F^{2}q_{z}^{2}}{\omega^{2}-v_F^{2}q_{z}^{2}+i0^{+}} e^{-\mathbf{q}_{\perp}^{2}\ell_{B}^{2}/2}\label{eq:PIVLLL}\\
\Pi^R_A\left(q\right) & \simeq\frac{\ell_{B}^{-2}}{2\pi^{2}\hbar v_F}\,\frac{v_F q_{z}\omega}{\omega^{2}-v_F^{2}q_{z}^{2}+i0^{+}} e^{-\mathbf{q}_{\perp}^{2}\ell_{B}^{2}/2}\label{eq:PIALLL}.
\end{align}
Unless the phonon momentum is nearly perpendicular to the magnetic field, Eqs.~(\ref{eq:PIVLLL}) and (\ref{eq:PIALLL}) are valid provided that $\omega_0 l_B/v_F < 1$ and ${\bf q}^2 l_B^2 \ll 1$.
\begin{comment}
The corresponding interband contributions to $\Pi_V$ decrease
as $\frac{\mathbf{q}^{2}}{n\ell_{B}^{-2}}$ and can be neglected
\end{comment}
When ${\bf q}\perp {\bf B}$, the chiral contributions to $\Pi_V$ and $\Pi_A$ vanish at any finite $\omega$, whereas
the interband contributions involving the nonchiral LLs remain finite.
However, being weakly dependent on $\omega$ in the strong field regime, the latter are unlikely to change our results appreciably.
%Unless the phonon momentum is nearly perpendicular to the magnetic field, 
%However, the contributions involving nonchiral LLs are likewise supressed at sufficiently strong magnetic fields (
%the additional condition $\ell_{B}^{-2}\gg {\bf q}_\perp^2$ 
%\mathbf{q}_{\perp}| |q_{z}\pm(\mu\pm b_0)|/(\hbar v_F)$ 
%is required in order to justify Eqs.~(\ref{eq:PIVLLL}) and (\ref{eq:PIALLL}).
%These conditions are discussed in more detail in Appendix B.

In the quantum limit, $\Pi_V^R$ and $\Pi_A^R$ are independent of $\mu$ and $b_0$.
Even though both response functions diverge at $\omega= v_F q_z$, a careful inspection evidences that the right hand side of Eq.~(\ref{eq:GenDisp}) remains finite at $\omega=v_F q_z$.
Therefore, there will not be a pole of the phonon spectral function for  $\omega\simeq v_F q_z$.
In contrast, the right hand side of Eq.~(\ref{eq:GenDisp}) diverges for the zeros of $\varepsilon(q)$; this implies the appearance of plasmon poles in the phonon spectral function.

Another observation from Eqs. (\ref{eq:PIVLLL}) and (\ref{eq:PIALLL}) is that $\Pi^R_V/\Pi^R_A = v_F q_z/\omega$. 
This relation, taken together with Eq.~(\ref{eq:bc}), has a significant physical consequence: in a strong magnetic field, Weyl fermions imprint a nonzero effective charge on ${\bf q}=0$ pseudoscalar optical phonons, but not on ${\bf q}=0$ scalar phonons.
Specifically, the induced charge at ${\bf q}\simeq 0$ reads
\begin{equation}
\label{eq:qst}
{\bf Q}_{\rm el}(\omega) \simeq i\frac{e^2}{\pi h} \frac{D_z a^2}{\hbar \omega} {\bf B},
\end{equation}
where $D_z$ is the optical deformation potential (in units of energy) produced by the pseudoscalar phonon and $a$ is the typical linear dimension of a unit cell. 
In the derivation of Eq.~(\ref{eq:qst}), we have used the long wavelength approximation $|g_z({\bf q})|^2 {\cal V}/M= D_z^2/(\rho a^2)$, where $\rho=N M/{\cal V}$ is the mass density of the crystal. 
It is interesting that ${\bf Q}_{\rm el}$ is directed along the magnetic field, that it is proportional to the quantum of conductance $e^2/h$, and inversely proportional to the frequency of the lattice vibration.
Although group theory can by itself predict an effective pseudoscalar phonon charge that is parallel to the magnetic field (regardless of whether the electronic structure hosts Weyl nodes or not), Eq.~(\ref{eq:qst}) provides a microscopic theory for this effect in WSM and links it to the chiral anomaly.
This finding confirms and generalizes the main result from Ref. [\onlinecite{Us}], where we predicted a magnetic-field-induced infrared activity as a signature of the chiral anomaly, to the strong magnetic field regime.
Roughly, the induced charge scales as $e D_z/(\hbar\omega)\Phi/\Phi_0$, where $\Phi$ is the magnetic flux traversing a cross section of the unit cell and $\Phi_0=h/e$ is the quantum of flux.  
For typical parameter values ($D_z=10 {\rm eV}$, $\hbar\omega=10 {\rm meV}$, $a=5 \AA$), the induced charge reaches $0.2 e$ (in absolute value) at a field of $10$ T.

Inserting Eqs.~(\ref{eq:PIVLLL}) and (\ref{eq:PIALLL}) in Eq.~(\ref{eq:GenDisp}),
we obtain the dispersions for purely scalar (``sc'') and purely pseudoscalar (``ax'') phonons:
\begin{widetext}
\begin{align}
\label{eq:sol_anal_1}
\omega_{\text{sc},\pm}^{2}({\bf q}) & =\frac{1}{2}\Biggl\{\omega_{{\rm ph},{\bf q}}^{2}+\omega_{{\rm pl},{\bf q}}^{2}\pm\Biggl[\left(\omega_{{\rm ph},{\bf q}}^{2}-\omega_{{\rm pl},{\bf q}}^{2}\right)^{2}+
\frac{2}{\pi^2}\omega_{{\rm pl},{\bf q}}^2\frac{|g_0 ({\bf q})|^2}{\hbar v_F l_B^2}\frac{e^{-{\bf q}_\perp^2 l_B^2/2}} {1 + \frac{2}{\pi}\frac{\alpha}{{\bf q}^2 l_B^2}e^{-{\bf q}_\perp^2 l_B^2/2}}\Biggr]^{\frac{1}{2}}\Biggr\}
\nonumber\\
\omega_{\text{ax},\pm}^{2} ({\bf q}) & =\frac{1}{2}\Biggl\{\omega_{{\rm ph},{\bf q}}^{2}+\omega_{{\rm pl},{\bf q}}^{2}\pm\Biggl[\left(\omega_{{\rm ph},{\bf q}}^{2}-\omega_{{\rm pl},{\bf q}}^{2}\right)^{2}+\frac{2}{\pi^2}\omega_{{\rm pl},{\bf q}}^2\frac{|g_z ({\bf q})|^2}{\hbar v_F l_B^2} e^{-{\bf q}_\perp^2 l_B^2/2}\Biggr]^{\frac{1}{2}}\Biggr\},
\end{align}
\end{widetext}
where $\omega_{{\rm ph},{\bf q}}\simeq \omega_0$ is a constant in the long wavelength regime of interest, 
\begin{equation}
\omega_{{\rm pl},{\bf q}}=v_F q_z\sqrt{1+\frac{2}{\pi} \frac{\alpha}{{\bf q}^2 l_B^2 } e^{-{\bf q}_\perp^2 l_B^2/2}}\label{eq:plasmonbare}
\end{equation}
is the bare plasmon frequency at momentum {\bf q} satisfying $\text{Re}\,\varepsilon({\bf q},\omega_{{\rm pl},{\bf q}})=0$, and 
\begin{equation}
\alpha = \frac{e^2}{4\pi \varepsilon_\infty \hbar v_F}
\end{equation}
is the effective fine structure constant in the WSM.
The plasmon frequency grows monotonically with $|{\bf q}|$ (provided that $\alpha<\pi$, which is expected for most WSM) and $B$.

In Eq.~(\ref{eq:sol_anal_1}), we have restricted ourselves to the real part of the frequency.
In fact, the contributions from $\Pi_V$ and $\Pi_A$ to the phonon linewidth are proportional to $\delta\left(\omega^2-v_F^2 q_z^2\right)$, and hence vanish everywhere along the dispersions in Eq.~(\ref{eq:sol_anal_1}).
Anharmonic phonon effects and the finite instrument resolution, captured by the parameter $\eta$, will be incorporated in the numerical results displayed below. 

In the absence of electron-phonon interactions, the solutions of
Eq.~(\ref{eq:sol_anal_1}) tend to $\omega_0$ and $\omega_{{\rm pl},{\bf q}}$, which correspond to decoupled optical phonons and plasmons. 
Electron-phonon interactions hybridize the plasmon and the optical phonon.
This hybridization, which can be attributed to an effective phonon charge induced by Weyl fermions, becomes notable when the resonance condition $\omega_{{\rm ph},{\bf q}}=\omega_{{\rm pl},{\bf q}}$ is satisfied. 
Far from resonance, the solutions in Eq.~(\ref{eq:sol_anal_1}) are $\omega_0$ and $\omega_{{\rm pl},{\bf q}}$, with small corrections of second order in the electron-phonon coupling.
At resonance, the phonon and the plasmon couple strongly and an avoided crossing (or ``anticrossing'') emerges between the two dispersion branches.
Even though there exists an extensive literature on the plasmon-phonon hybridization in polar semiconductors, both in the absence and in the presence of magnetic fields,\cite{Mahan,CardonaLSS,Falkovsky,Wysmolek} 
in WSM the hybridization at long wavelengths is induced by magnetic fields even when the material is intrinsically nonpolar.
%{\bf possibly the effect exists in non WSM}
As we elaborate below, the hybridization gap in this case is about an order of magnitude smaller than in polar semiconductors like GaAs, though it may reach observability at high magnetic fields.
%, and that the induced Born charge is linear in the field at strong fields.
Another peculiar feature of Eq.~(\ref{eq:sol_anal_1}) is that the plasmon-phonon hybridization is turned off at ${\bf q}=0$ for scalar phonons, but not for the pseudoscalar phonons.

\begin{figure}[H]
\includegraphics[width=0.78\columnwidth]{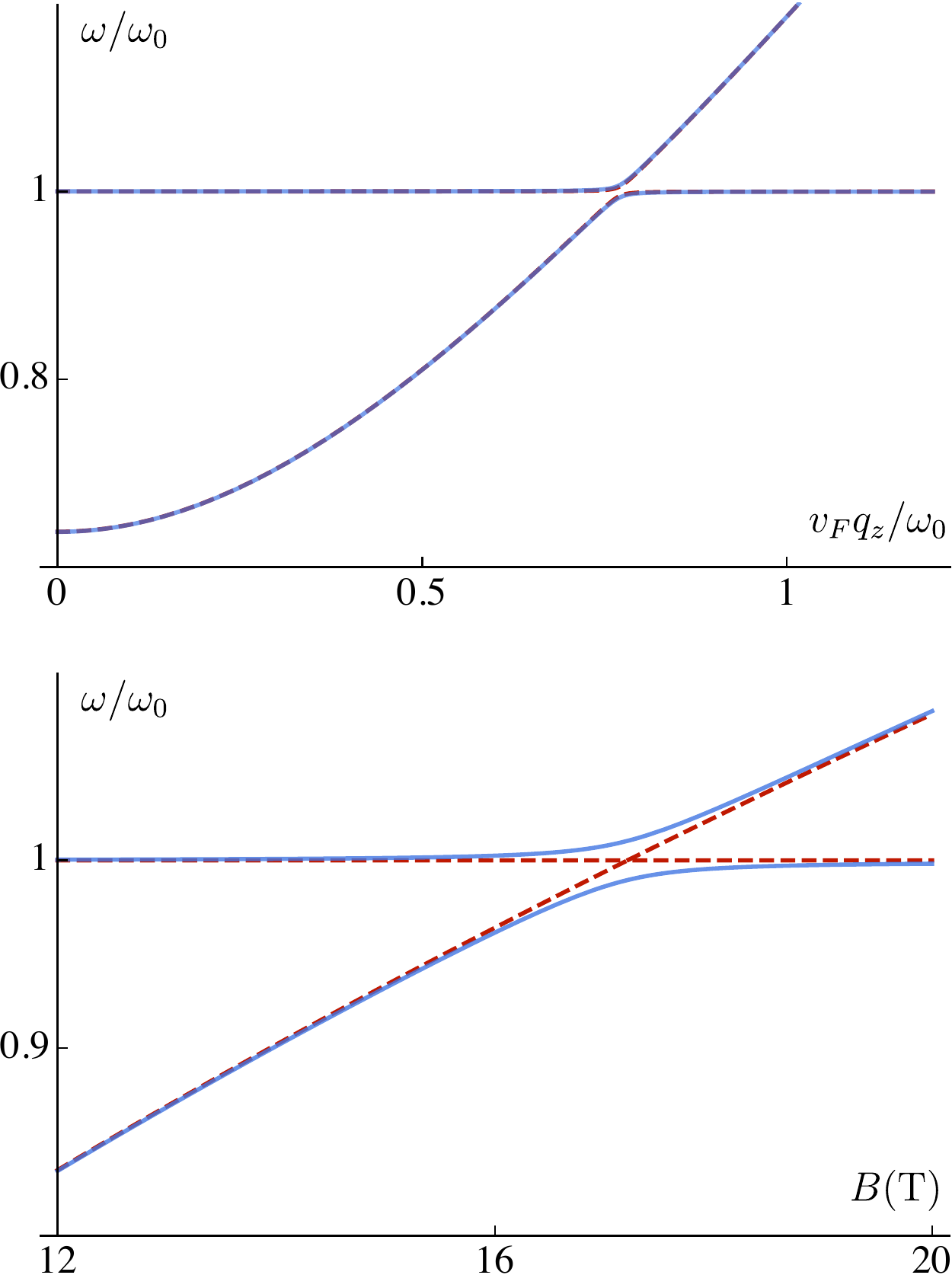}
\caption{Coupled phonon and plasmon dispersion relation in the quantum limit, obtained from Eq.~(\ref{eq:sol_anal_1}). 
  The solid (dashed) lines correspond to the case of pseudoscalar (scalar) phonons, whose deformation potentials are assumed to be equal for illustrative purpose.
(Top) Dispersion as a function of the momentum, for a fixed magnetic field of $B=7 {\rm T}$.
(Bottom) Dispersion as a function of the magnetic field, for a (small) fixed momentum $\hbar v_F |{\bf q}| = 0.1 {\rm meV}$.
  The parameters common to both panels are: $\theta=0$ for the angle between the magnetic field and the phonon momentum, %$\mu=0$ for the chemical potential,
  $\hbar\omega_0= 10 {\rm meV}$ for the bare optical phonon energy, $D_{0 (z)}=10\text{eV}$ for the scalar and pseudoscalar deformation potentials, $a=4\AA$ for the lattice constant, $\rho=10^4 {\rm kg/m}^3$ for the mass density, $\varepsilon_{\infty}=30\varepsilon_{0}$ for the permittivity from high-energy electronic bands, and $v_F=2\times 10^{5}\text{m/s}$ for the Fermi velocity.
\label{fig:disp1}}
\end{figure}

Figure \ref{fig:disp1} displays $\omega_{\rm sc}$ and $\omega_{\rm ax}$ for ${\bf q} || {\bf B}$.
The panel on top shows the phonon and plasmon dispersions for a fixed magnetic field: the plasmon-phonon anticrossing occurs both for pseudoscalar and scalar phonons.
The panel at the bottom shows the phonon and plasmon frequencies at zero momentum as a function of the magnetic field strength: 
in this case, there exists a phonon-plasmon hybridization gap only for pseudoscalar phonons.
This is a signature of the VVA diagram and the chiral anomaly.
%, and can be connected to a magnetic-field-induced Born charge ({\bf cite}).

For the remainder of this subsection, we discuss the avoided crossing more quantitatively.
We begin by establishing the conditions under which it takes place, and follow
by estimating its  magnitude.

As mentioned above, the anticrossing occurs when $\omega_0=\omega_{{\rm pl},{\bf q}}$, i.e. when 
\begin{equation}
\label{eq:cond}
v_F^2 \cos^2\theta \left({\bf q}^2 + \frac{2\alpha e B}{\pi\hbar} e^{-{\bf q}^2 \sin^2\theta \,l_B^2/2}\right) = \omega_0^2,
\end{equation}
where $\theta=\cos^{-1}(q_z/|{\bf q}|)$ is the angle between the magnetic field and the phonon momentum. 
For simplicity, hereafter we approximate $\exp(-{\bf q}^2 \sin^2\theta\, l_B^2/2) \simeq 1$.
This is justified in the regime of strong magnetic fields and long wavelength, regardless of the value of $\theta$.

For a fixed momentum ${\bf q}$, Eq.~(\ref{eq:cond}) gives the value of the magnetic field at which the avoided crossing takes place.
Conversely, given $B$ and $\theta$, Eq.~(\ref{eq:cond}) determines the value of $|{\bf q}|$ at which the avoided crossing takes place.
Clearly, as ${\bf q}$ becomes misaligned with ${\bf B}$, a stronger magnetic field is required in order to attain resonance.
In the limit of ${\bf q}\perp {\bf B}$ (i.e. $\cos\theta=0$), the necessary magnetic field tends to infinity, i.e. the avoided crossing is no longer possible. 
This is due to the fact that the plasmon frequency in Eq.~(\ref{eq:plasmonbare}) vanishes when $\cos\theta \to 0$.

Equation~(\ref{eq:cond}) imposes a maximum value of the magnetic field $B_{\rm max}$ for which the anticrossing condition can be achieved:
\begin{equation}
\label{eq:Bmax}
B_{\rm max}=\frac{\pi}{2\alpha}\frac{\hbar}{e}\frac{\omega_0^{2}}{v_F^{2}\cos^{2}\!\theta}.
\end{equation}
This is the field required to bring the plasmon and the phonon into resonance at $|{\bf q}|\simeq 0$. 
If $B>B_{\rm max}$, the plasmon frequency exceeds the phonon frequency for all momenta, hence preventing a plasmon-phonon resonance.
For $v_F=2\times 10^{5}{\rm m/s}$, $\varepsilon_{\infty}=40\varepsilon_{0}$ and $\hbar\omega_0=15\text{meV}$, we have $B_{\rm max}[T]\simeq 50/\cos^{2}\!\theta$, which can be made as large as desired by an appropriate choice of $\theta$.

On a related note, the magnetic field appearing in Eq.~(\ref{eq:cond}) needs to be larger than $B_{QL}$; otherwise, Eqs.~(\ref{eq:sol_anal_1}) and (\ref{eq:plasmonbare})  would not be valid to begin with.
%This imposes a maximum value of $|{\bf q}|$ for which the phonon-plasmon anticrossing can occur:
%\begin{equation}
%\label{eq:qcond}
%|{\bf q}|_{\rm max}= \sqrt{\frac{2\alpha}{\pi}\frac{e}{\hbar} (B_{\rm max}-B_{QL})}.
%\end{equation}
%If $|{\bf q}|>|{\bf q}|_{\rm max}$, the plasmon frequency exceeds the phonon frequency for all magnetic fields in the quantum limit, thereby precluding a resonance. 
%Likewise, there is no avoided crossing in the quantum limit at any ${\bf q}$ if the argument of the square root in Eq.~(\ref{eq:qcond}) is negative.
As we shall see in Sec. \ref{subsec:Intermediate-field}, avoided crossings can also take place away from the quantum limit ($B<B_{QL}$).
Yet, in that case, the contributions from the nonchiral LL to $\Pi_V$ and $\Pi_A$ can no longer be ignored.
\begin{comment}
\[
\frac{\pi}{2\tilde{\alpha}}\frac{\hbar}{e}\frac{\omega_0^{2}}{v^{2}\cos^{2}\!\theta}=\frac{2\pi^{2}\varepsilon_{\infty}}{e^{3}}\frac{\hbar^{2}}{v}\frac{\omega_0^{2}}{\cos^{2}\!\theta}\sim\frac{20\times3.10^{-10}10^{-68}4.10^{26}}{5.10^{-57}10^{5}\cos^{2}\!\theta}=\frac{48\text{T}}{\cos^{2}\!\theta}
\]
\end{comment}

Having listed the conditions for the avoided crossing, we now estimate its magnitude from Eq.~(\ref{eq:sol_anal_1}) by evaluating the hybridization gap $\Delta\omega\equiv \omega_{+} - \omega_{-}$ at $\omega_0=\omega_{{\rm pl},{\bf q}}$, both for scalar and pseudoscalar phonons.
We get
\begin{align}
\label{eq:gap}
\hbar \Delta\omega_{\rm sc} &\simeq \frac{D_0}{\pi a}\sqrt{\frac{e B}{2\rho v_F}\left(1-\frac{B}{B_{\rm max}}\right)}\nonumber\\
\hbar \Delta\omega_{\rm ax} &\simeq \frac{D_z}{\pi a}\sqrt{\frac{e B}{2\rho v_F}},
\end{align}
where $B$ satisfies Eq.~(\ref{eq:cond}) and $D_0$ is the optical deformation potential produced by the scalar phonon ($D_z$, $a$ and $\rho$ have been defined below Eq.~(\ref{eq:qst})).
In the derivation of Eq.~(\ref{eq:gap}), we have once again used the long wavelength approximation $|g_{0(z)}({\bf q})|^2{\cal V}/M= D_{0 (z)}^2/(\rho a^2)$, with a single lattice constant.
The fact that $\Delta\omega$ scales linearly with $|g_0|$ or $|g_z|$ evidences the non-perturbative effect of the electron-phonon interaction when the bare plasmon and phonon come into resonance.

Equation (\ref{eq:gap}) states that, for a given deformation potential ($D_0=D_z$), $\Delta\omega_{\rm ax}>\Delta\omega_{\rm sc}$. 
This is particularly true when ${\bf q}\to 0$, for which Eq.~(\ref{eq:cond}) dictates $B\to B_{\rm max}$ and thus $\Delta\omega_{\rm sc}\to 0$.
In contrast, $\Delta\omega_{\text{ax}}\neq 0$  at ${\bf q}\to 0$.
As mentioned above, this difference in behavior originates from the electronically induced phonon effective charge at ${\bf q}=0$ (cf. Eq.~(\ref{eq:qst})), which is nonzero (zero) for pseudoscalar (scalar) phonons.
At any rate, the ${\bf q}\to 0$ avoided crossing can be realized only in WSM with sizeable electron-electron interactions.
In the noninteracting limit ($\alpha\to 0$), $B_{\rm max}$ becomes unattainably large.
In this limit, the plasmon dispersion merges with the pole of $\Pi_A$, the avoided crossing in the phonon dispersion takes place only at $|{\bf q}| = \omega_0/(v_F \cos\theta)$, and $\Delta\omega_{\rm ax}=\Delta\omega_{\rm sc}$ (for $D_0=D_z$).
%a given deformation potential.

\begin{figure}
\includegraphics[width=\columnwidth]{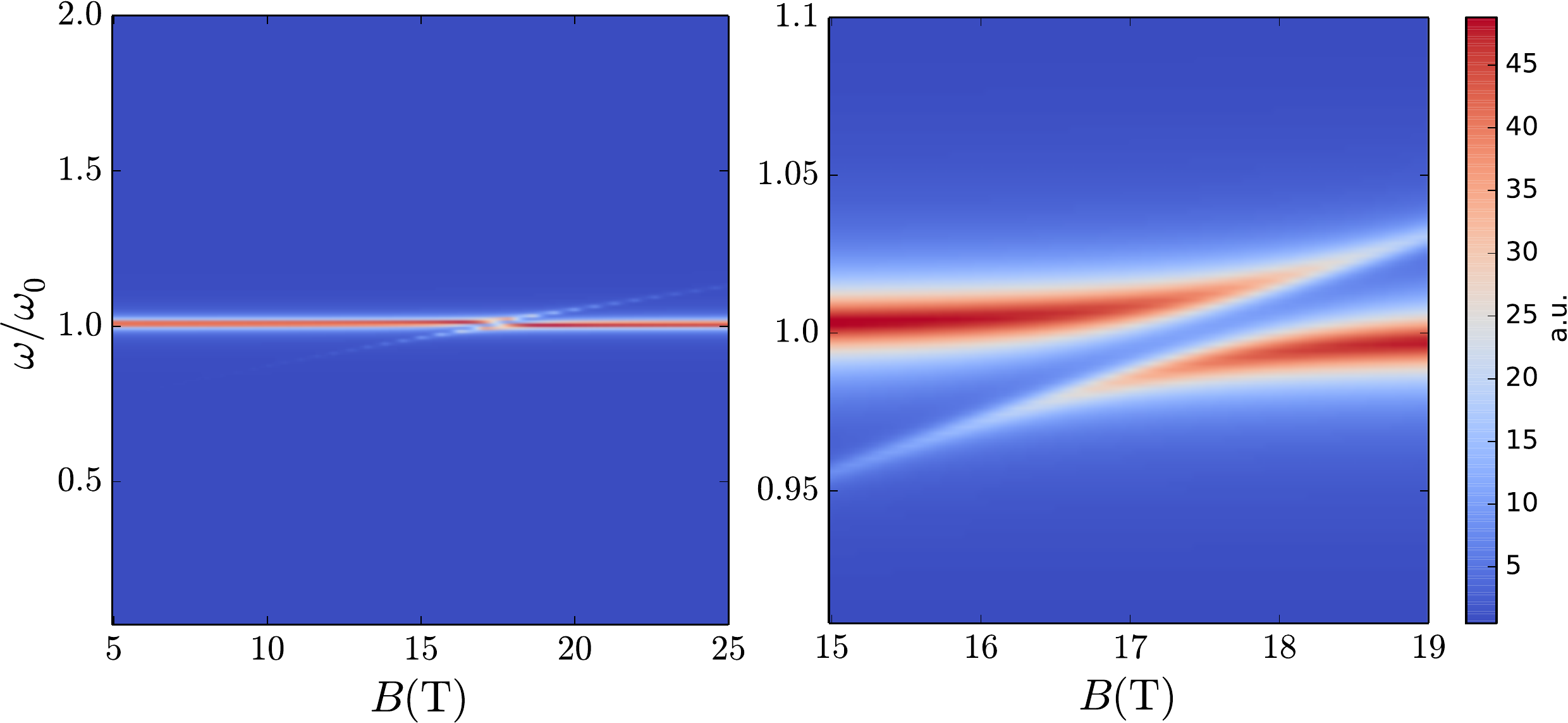}
\caption{Optical phonon spectral function evaluated from Eq.~(\ref{eq:spectral}) as a function of the magnetic field, in the quantum limit. 
The right panel zooms in on the anticrossing region of the left panel. 
The parameters chosen are $\cos\theta=0.8$, $\hbar v_F |{\bf q}|=8 {\rm meV}$ and $\eta=0.1 {\rm meV}$. The numerical values of the remaining parameters coincide with those of Fig.~\ref{fig:disp1}.
}
\label{fig:spec_lin}
\end{figure}

One strategy to experimentally observe the plasmon-phonon anticrossing is to measure the frequency of a phonon as a function of the magnetic field, at fixed phonon momentum. 
Let us denote as $B^*$ the maximum magnetic field attainable in the laboratory, and let us suppose that $B_{QL}<B^*<B_{\rm max}/2$. 
Then, it is advantageous to fix the phonon momentum to the value that satisfies Eq.~(\ref{eq:cond}) with $B\lesssim B^*$; this choice will lead to the maximal hybridization gap (given by Eq.~(\ref{eq:gap}) with $B$ replaced by $B^*$).  
The outcome of the measurement should resemble Fig.~\ref{fig:spec_lin}.
This figure displays the phonon spectral function, given by the imaginary part of
\begin{equation}
\label{eq:spectral}
D({\bf q},\omega) = \frac{2 \omega_{\bf q}}{(\omega+ i \eta)^2 - R({\bf q},\omega)},
\end{equation}
where $R({\bf q},\omega)$ equals the right hand side of Eq.~(\ref{eq:GenDisp}).

In order to make the avoided crossing observable, it is necessary that $\Delta\omega >\eta$, where $\eta$ is determined by anharmonic phonon effects and the finite spectral resolution of the experiment (we recall that the imaginary parts of the retarded response functions practically vanish in the quantum limit).
A common lower bound for $\hbar\eta$ is $\simeq 0.1 {\rm meV}$ (spectral resolution of $\simeq 1 {\rm cm}^{-1}$), which can be realized at low temperatures and with high precision probes (e.g. Raman spectroscopy). 
%({\bf add refs}).
For typical material parameters ($\rho=10^{4}{\rm kg/m}^3$, $a=5\mathring{\text{A}}$,
$v_F=2\times10^{5}{\rm m/s}$, and $D=5\text{eV}$), we get $\hbar \Delta\omega\simeq 30 \sqrt{B^*[\text{T}]}\mu\text{eV}$.
This value is independent of the strength of electron-electron interactions.
Thus, fields of about $10 {\rm T}$ are required just to match the gap with the limit of experimental resolution. 
The situation becomes more favorable if the WSM contains $N_W>1$ pairs of degenerate Weyl nodes, in which case $\hbar\Delta\omega\simeq 30 \sqrt{N_W B^*[{\rm T}]} \mu\text{eV}$.
Incidentally, when $N_W>1$, Eqs.~(\ref{eq:cond}) and (\ref{eq:Bmax}) must be modified via $B \to N_W B$.
%{\bf OK? Need to redo calculations when $\Pi \to N_W \Pi$} \textcolor{green}{OK. In the quantum limit, $B$ appears only as a factor in front of the polarization functions. Since we have to multiply those polarization function by $N_W$, we just have to multiply $B$ by $N_W$ in the final results.}

The preceding estimates show that the magnetic-field-induced plasmon-phonon hybridization may be observable in high precision measurements conducted at low temperature and high magnetic fields.
At any rate, given the sensitivity of our estimates to the unit cell dimensions, the Fermi velocity, and the deformation potential, it would be highly desirable to determine them for real WSM from first-principles electronic structure calculations.

\subsection{Intermediate magnetic fields\label{subsec:Intermediate-field}}

In moderately doped WSM (or, alternatively, in WSM where the nodes of opposite chirality are located at largely different energies), the highest attainable magnetic fields may not suffice to reach the quantum limit.
For instance, $B_{QL}\simeq 50 {\rm T}$ for $|\mu\pm b_0|\simeq 50 {\rm meV}$. 
In this subsection, we will determine the optical phonon dispersion when $B\lesssim B_{\rm QL}$ (``intermediate'' field regime), where the Fermi level intersects a few nonchiral LLs in addition to the chiral LLs.
\cite{interm}

Equation ~(\ref{eq:GenDisp}) is valid for any magnetic field.
However, Eqs.~(\ref{eq:PIVLLL}) and (\ref{eq:PIALLL}) no longer hold at $B<B_{QL}$, because the contribution from nonchiral Landau levels to  $\Pi_V$ and $\Pi_A$ must be added.
One influence of the nonchiral LLs is that the phonon dispersion becomes dependent on both $b_0$ and $\mu$.  
Moreover, nonchiral LLs lead to a finite imaginary part of $\Pi_V$ and $\Pi_A$ for regions of nonzero area in the $(\omega,{\bf q})$ space, thereby creating extended electron-hole continua in the phonon spectral function. 
Inside these continua, the phonon frequency is damped proportionally to the square of the electron-phonon coupling.

We will begin by discussing the situation with ${\bf q}||{\bf B}$ and $|b_0|\ll |\mu|$, where two simplifications take place.
First, the contribution from nonchiral LLs to $\Pi_A$ vanishes, so that Eq.~(\ref{eq:PIALLL}) retains its validity.
The underlying reason is that the nonchiral contributions from the two nodes cancel each other.
Second, the only interband transitions contributing to $\Pi_V$ are $-n \to n$. 
Consequently, each pair $(-n,n)$ of nonchiral LLs contributes to $\Pi_V$ independently as if it were a one-dimensional Dirac fermion with mass $\Delta_n\equiv\sqrt{2 |n|}\hbar v_F\ell_{B}^{-1}$.
This mapping allows to use analytical expressions available in the literature\cite{Thakur} and to benchmark our numerical calculations. 

\begin{figure}
\includegraphics[width=1\columnwidth]{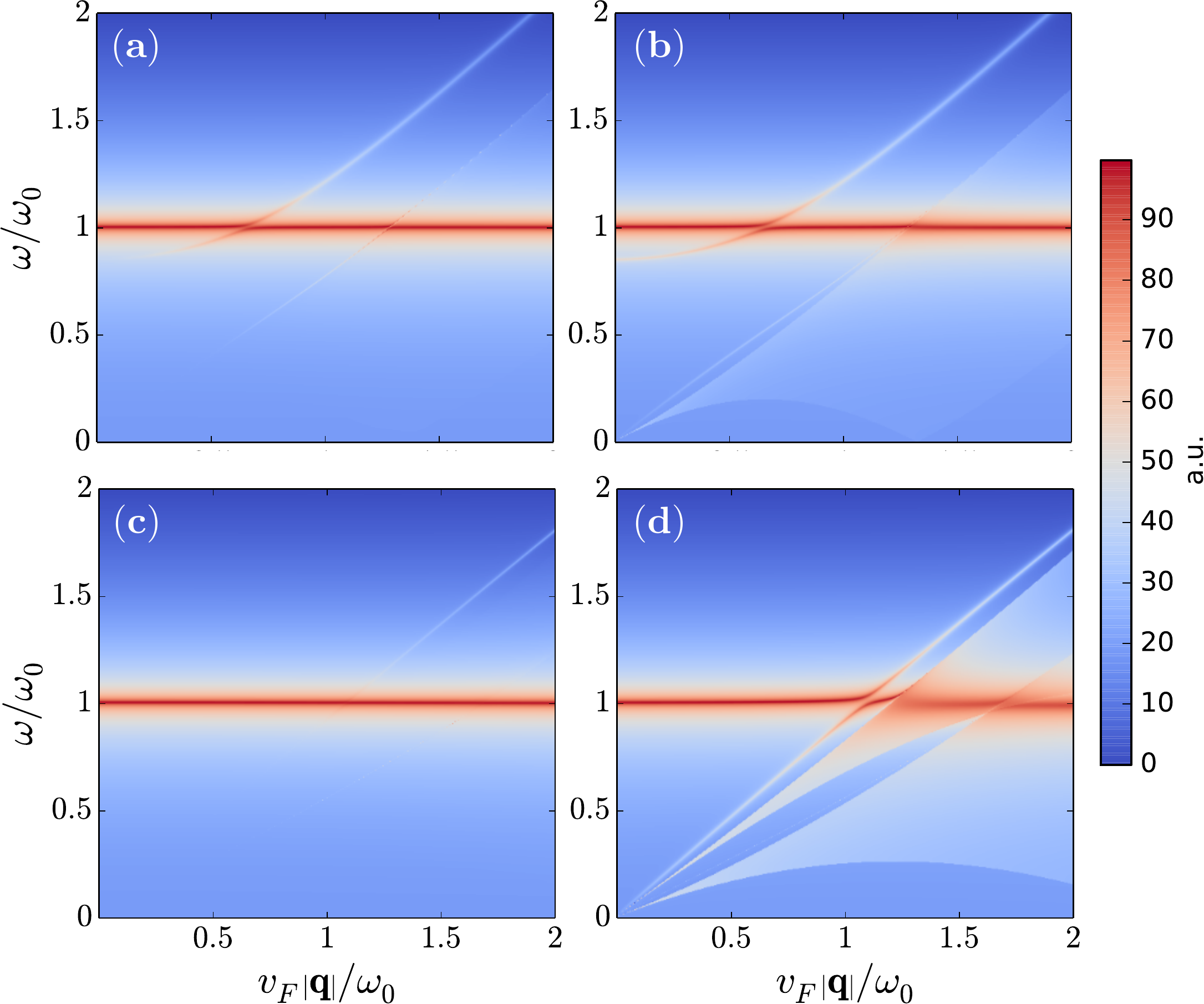}\caption{
Optical phonon spectral function in the intermediate magnetic field regime, evaluated from Eq.~(\ref{eq:spectral}) and represented in logarithmic scale. 
The parameter values are: $\theta=b_0=0$, $\hbar\omega_0=10{\rm meV}$, $\rho=10^4 {\rm kg/m}^3$, $D_{0(z)}=10{\rm eV}$, and $\eta=0.1{\rm meV}$. 
Panels (a) and (b) display a plasmon-phonon anticrossing in the cases of purely scalar and purely pseudoscalar electron-phonon interactions, respectively. 
In both panels, $B=8\text{T}$ and $\mu=\hbar\omega_0$.
In panels (c) and (d), the chemical potential and the magnetic field are sufficiently high ($\mu=3\hbar\omega_0$ and $B=30{\rm T}$) that the plasmon frequency is pushed well above the phonon frequency, thereby preventing their resonance. 
In spite of this, panel (d) (purely pseudoscalar electron-phonon interactions) contains a large anticrossing between the optical phonon and particle-hole excitations originating from nonchiral LLs.
This anticrossing is washed out by electronic screening in the case of purely scalar electron-phonon interactions (panel (c)). 
\label{fig:Phonon-spectral-density}}
\end{figure}

\begin{figure}
\includegraphics[width=1\columnwidth]{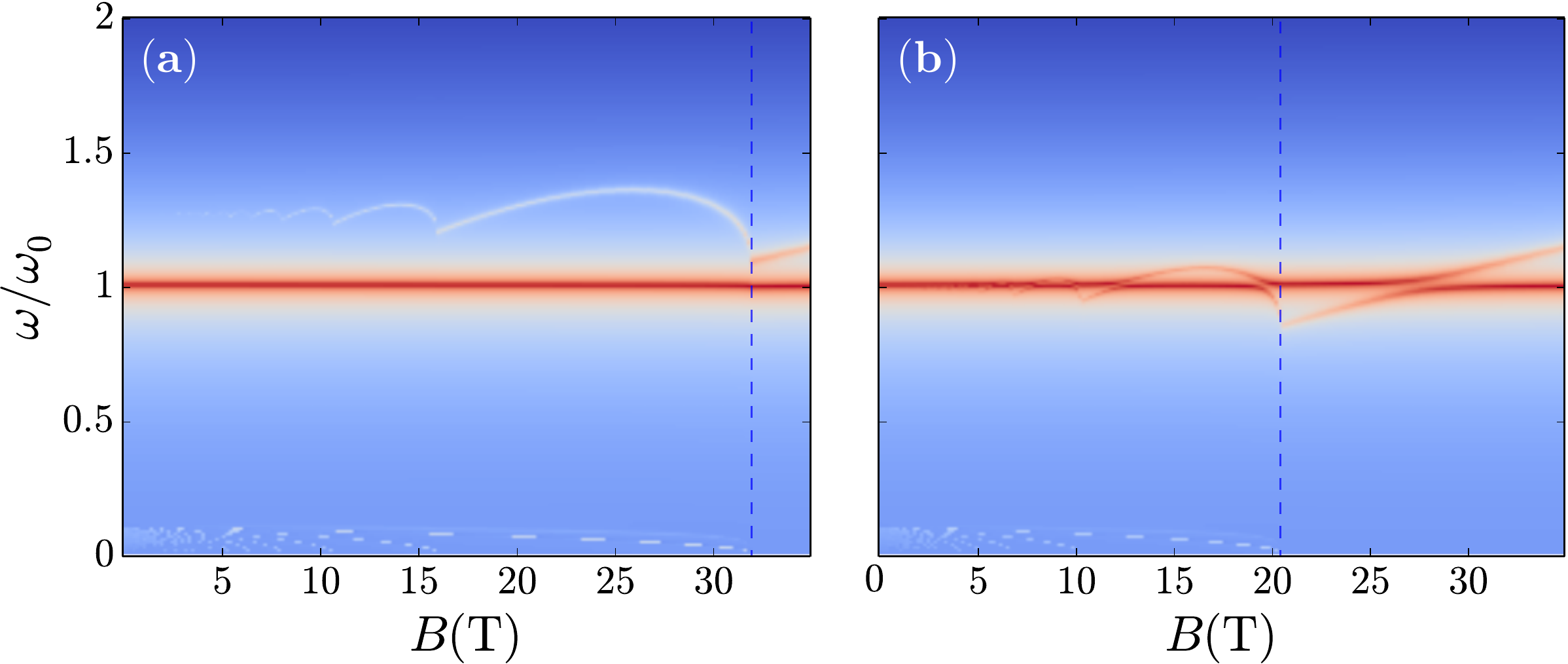}\caption{
Optical phonon spectral function in the intermediate and strong magnetic field regimes, evaluated from Eq.~(\ref{eq:spectral}) and represented in logarithmic scale. 
The vertical dashed lines correspond to $B=B_{QL}$. The color code is the same as in the previous figure.
The parameter values are: $\theta=b_0=0$, $\hbar\omega_0=10{\rm meV}$,  $\hbar v_F |{\bf q}| = 1 {\rm meV}$,  $\rho=10^4 {\rm kg/m}^3$, $D_0=2{\rm eV}$, $D_z=8{\rm eV}$ and $\eta=0.1{\rm meV}$. 
In panel (b), the chemical potential is fine tuned ($\mu=1.65 \hbar\omega_0$) so that the optical phonon shows large quantum oscillations due to its strong coupling to the plasmon.
Panel (a) displays the situation with $\mu=2\hbar\omega_0$, in which quantum oscillations of the phonon frequency are unobservable because the phonon is largely decoupled from the plasmon (the latter is responsible for the faint traces in the phonon spectral function at $\omega>\omega_0$).
\label{fig:Phonon-spectral-density-1}}
\end{figure}

The phonon spectral function at $\theta=b_0=0$ is depicted in Figs.~\ref{fig:Phonon-spectral-density} and \ref{fig:Phonon-spectral-density-1}.
Figures \ref{fig:Phonon-spectral-density}(a) and \ref{fig:Phonon-spectral-density}(b) confirm that the phonon-plasmon anticrossing can also take place away from the quantum limit.
As a rule of thumb, in order to find such an avoided crossing at $B\lesssim B_{QL}$, $\omega_0$ must be larger than the plasmon frequency at zero field and zero momentum, i.e. we need\cite{Zhou}
\begin{equation}
\hbar \omega_0 \gtrsim \sqrt{\alpha}|\mu\pm b_0|.
\end{equation} 
Then, as $|{\bf q}|$ is increased, the plasmon frequency grows monotonically until it gets in resonance with the optical phonon. 

If  $|\mu|\gg\omega_0$, the intraband part of $\Pi_V$ pushes the plasmon dispersion far above the phonon dispersion for all momenta, and the plasmon-phonon hybridization becomes negligible.
Nevertheless, we find nontrivial electronic fingerprints in the phonon spectrum even when the phonon and the plasmon are decoupled.
For instance, Fig.~\ref{fig:Phonon-spectral-density}(d) features an avoided crossing just above the electron-hole continuum.
This anticrossing is not associated with the plasmon; rather, it originates from logarithmic divergences in $\Pi_V$ that are located at frequencies\cite{sing} 
\begin{equation}
\omega_{n,{\bf q}}=v_F q_{z}\left[\sqrt{1-\frac{\Delta_{n}^{2}}{\mu^{2}}}+\mathcal{O}\left(\frac{\hbar v_F q_{z}}{\mu}\right)\right].\label{eq:logpeakpos}
\end{equation}
%{\bf Proof? What are the values of $n$? Does the optical phonon get overdamped at larger momenta?} 
These divergences arise from intraband transitions within nonchiral LLs. 
As such, they disappear in the quantum limit, due to Pauli blocking.
For pseudoscalar phonons, a divergence of $\Pi_V$ implies a peak in the spectral function at $\omega\simeq\omega_{n,{\bf q}}$ because the correction to the phonon dispersion in Eq.~(\ref{eq:GenDisp}) contains $\Pi_V$ without any screening factor.
Anticrossings with the optical phonon occur when $\omega_{n,{\bf q}} = \omega_0$, for all values of $n$ such that $\Delta_n<|\mu|$.
Thus, the number of anticrossings equals the number of nonchiral LLs intersecting the Fermi energy (note that for the parameters of Fig.~\ref{fig:Phonon-spectral-density}(d), there is a single nonchiral LL that intersects the Fermi energy).
As shown in Fig. \ref{fig:Phonon-spectral-density}(c), the anticrossing is suppressed for scalar phonons.
The reason is that, for a scalar phonon, the electronic contribution to the phonon dispersion goes as $\Pi_V/\varepsilon$, which 
remains finite when $\Pi_V$ diverges.

Another characteristic aspect of the intermediate magnetic field regime is the emergence of quantum oscillations in the frequency of the optical phonon, inherited from its coupling to electrons.
For typical parameters, we find that the amplitude of quantum oscillations in the optical phonon frequency is smaller than $\eta$, thereby precluding their observability. 
An exception takes place when $\omega_{\rm p}(B=0) \simeq \omega_0$, i.e. when $\hbar\omega_0\simeq |\mu| \sqrt{\alpha/\pi}$.
In this case, the optical phonon and the plasmon are close to resonance for a range of values of $B$, and thus the magnetic oscillations in the optical phonon frequency become sizeable.
%because due to its strong coupling to the plasmon each time the plasmon frequency crosses $\omega_0$. 
This situation is reflected in Fig.~\ref{fig:Phonon-spectral-density-1}.
Unfortunately, since neither $\omega_0$ nor $\mu$ are tunable in WSM, it appears unlikely that quantum oscillations will be seen for optical phonons.

\begin{figure}
\includegraphics[width=1\columnwidth]{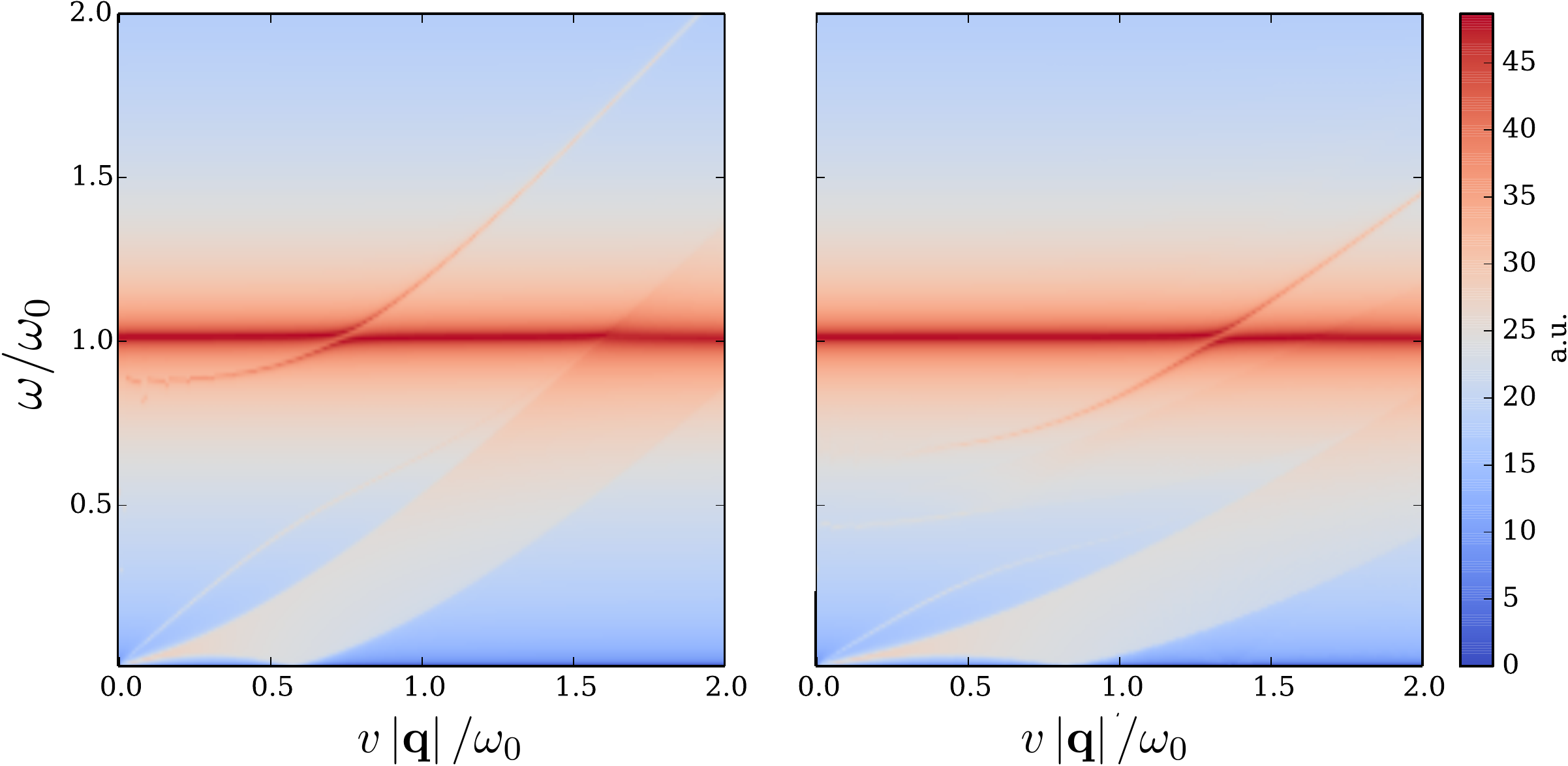}\caption{
Optical phonon spectral function in the intermediate magnetic field regime, evaluated from Eq.~(\ref{eq:spectral}) and represented in logarithmic scale. 
The parameter values are: $b_0=0$, $B=5{\rm T}$, $\hbar\omega_0=10{\rm meV}$, $\mu=1 \hbar\omega_0$, $\rho=10^4 {\rm kg/m}^3$, $D_0=0$, $D_z=10{\rm eV}$ and $\eta=0.1{\rm meV}$. 
The relative angle between the magnetic field and the phonon momentum is $\theta=0$ in the left panel, and $\theta=0.8$ in the right panel.
For smaller values of $|\cos\theta|$ and fixed $B$, the phonon-plasmon anticrossing moves to higher momenta.} 
\label{fig:inter_theta}
\end{figure}

Thus far we have treated the case $\theta=b_0=0$. 
For $\theta\neq 0$, $\Pi_V$ and $\Pi_A$ must be computed numerically from Eq.~(\ref{eq:pill}) (see also Appendix B).
Taking $\theta\neq 0$ brings about two changes. 
First, the plasmon frequency decreases monotonously as $\cos\theta$ decreases, much like in the quantum limit.
Consequently, the plasmon-phonon anticrossing takes place at higher momenta, as shown in Fig.~\ref{fig:inter_theta}.
Second, interband transitions of the type $n\to n'$ with $n'\neq -n$ are no longer forbidden (though they are smaller than $-n \to n$ transitions by a factor $\left(\textbf{q}_\perp \ell_B\right)^{2||n|-|n'||}$, as can be verified from the expressions of Appendix B). 
These new transitions produce additional electron-hole continua, which further damp the optical phonon if they overlap with it.
In regard to $b_0\neq 0$, its main qualitative effect (not shown) is to split each electron-hole continuum into two, shifted in frequency with respect to one another by $b_0$.

\section{Acoustic phonons}
\label{sec:acoustic}

In the previous section, we have identified the magnetic-field-induced hybridization between plasmons and optical phonons as a potentially distinctive signature of Weyl fermions. 
This hybridization requires that the bare phonon frequency become equal to the bare plasmon frequency for a certain value of the magnetic field or the phonon momentum.
Such condition is hardly achievable in the case of long-wavelength acoustic phonons, because the sound velocity $c_s$ is much smaller than the Fermi velocity.
Notably, in the quantum limit (where the phonon-plasmon hybridization is strongest), $c_{s}\left|\mathbf{q}\right|=\omega_{\text{p}}\left(\mathbf{q}\right)$ is satisfied only if $|\cos\theta|<c_s/v_F\simeq 10^{-2}$. 
Current experiments are not able to tune $\theta$ with such precision.
%{\bf Does $\Pi_A=0$ in the static limit imply that sound velocity contains no information on chiral anomaly? This seems to contradict Spivak}

In this section, we present a theory on how Weyl fermions alter the sound velocity in the presence of a magnetic field.
This theory is motivated in part by recent~\cite{Ramshaw} and ongoing~\cite{Jeffrey} ultrasound experiments in TaAs.
As we show, the change in sound velocity under the application of a magnetic field can reflect the presence of
the chiral Landau levels and can display quantum oscillations which are particular to Weyl fermions.
A recent theory\cite{Andreev} has investigated the sound attenuation in WSM using the semiclassical approximation, but there are no predictions for the sound velocity itself.

The literature on how magnetic fields contribute to sound velocity in conducting materials is extensive.\cite{Luthi}
Here, we will follow the formalism developed in Sec. \ref{sec:phodis}. 
For concreteness, we will concentrate on scalar phonons and piezoelectric electron-phonon coupling, which are believed to be relevant in current ultrasound experiments.
The sound velocity in an insulating piezoelectric crystal is given by \cite{Carleton}
%contains contributions from the elastic force tensor
%and the piezoelectricity and can be written
\begin{equation}
c_{s}=\sqrt{c_0^{2}+\frac{1}{\rho}\frac{d^{2}}{\varepsilon_{\infty}}},\label{eq:noelSV}
\end{equation}
where $c_0$ is the sound velocity in the absence of piezoelectricity and $d$ is the piezoelectric coupling (whose value depends on the polarization of the acoustic wave, as well as on its direction of propagation).
%assumed for simplicity to be a constant).

In the presence of conduction electrons, the sound velocity is no longer $c_s$. 
Starting from Eq.~(\ref{eq:GenDisp}) and using $|g_0|^2 {\cal V}/M=(e d/\varepsilon_{\infty})^2/\rho$,\cite{Cardona} we obtain
\begin{equation}
  \omega^{2}\simeq\mathbf{q}^{2}\left(c_0^{2}+\frac{1}{\rho}\frac{d^{2}}{\varepsilon_{\infty}}\frac{1}{\varepsilon\left(\mathbf{q},\omega\right)}\right).
  \label{eq:dispAC_piezo}
\end{equation}
A formally identical equation has been obtained at zero magnetic field from a different approach in Refs.~[\onlinecite{Carleton, Hutson}]; the link becomes clear through the relation $\varepsilon({\bf q},\omega) = 1 + i {\bf q}\cdot\boldsymbol{\sigma}\cdot{\bf q}/(\omega |{\bf q}|^2 \varepsilon_\infty)$, 
%which can be obtained from the law of charge conservation.
where ${\boldsymbol\sigma}$ is the electrical conductivity tensor and we have used the fact that the electric field generated by the sound wave is longitudinal. 
%The expression from Refs.~[\onlinecite{Carleton, Huttson}] holds when ${\boldsymbol\sigma}$ is diagonal and isotropic.
%Yet, 
The larger the electrical conductivity of the sample is, the smaller the sound velocity becomes. 
At nonzero magnetic field, the tensor structure of the conductivity becomes nontrivial even in crystals with cubic symmetry.
This is particularly true for a WSM deep in the quantum limit, where $\sigma_{xx}=\sigma_{yy}=0$ (because the chiral LLs have no group velocity in the $x$ and $y$ directions) and $\sigma_{zz}\neq 0 \neq \sigma_{xy}=-\sigma_{yx}$.
In this situation,  $\varepsilon({\bf q},\omega) = 1+ i \sigma_{zz}({\bf q},\omega) \cos^2\theta/ (\varepsilon_\infty \omega)$.
%This expression illustrates the connection between the sound velocity and the chiral anomaly.

Because long wavelength acoustic phonons have frequencies that are small compared to the impurity scattering rate $\Gamma$, the effect of a finite electronic lifetime must be considered in the dielectric function of Eq.~(\ref{eq:dispAC_piezo}).
The simplest approach is to replace $\omega$ by $\omega+i\Gamma$ in $\Pi_V^R$, but this violates the particle number conservation.
A better (number conserving) approach is to use the expression\cite{Mermin, Ashcroft}
\begin{equation}
\Pi_V^R(\mathbf{q},\omega)=\frac{\left(\omega+i\Gamma\right)\Pi^R_{V0}(\mathbf{q},\omega+i\Gamma)\Pi^R_{V0}(\mathbf{q},0)}{\omega\Pi^R_{V0}(\mathbf{q},0)+i\Gamma\Pi^R_{V0}(\mathbf{q},\omega+i\Gamma)},\label{eq:EQMERMIN}
\end{equation}
where $\Pi^R_{V0}(\mathbf{q},\omega)$ is the retarded response function in the absence of disorder.
When $\omega\gg \Gamma$, Eq.~(\ref{eq:EQMERMIN}) gives approximately the same results as the number non-conserving approximation. 
%However, the regime pf  not relevant to long-wavelength acoustic phonons. 
In the following, we consider acoustic phonons for which $\Gamma\gg v_F |{\bf q}| \gg \omega \simeq c_s |{\bf q}|$ and ${\bf q}^2 l_B^2\ll 1$.
In this regime, the outcome from Eq.~(\ref{eq:EQMERMIN}) differs greatly from the number non-conserving approximation.
At the same time, we assume $\hbar v_F/l_B \gg \Gamma$ ($\Gamma < 5 \sqrt{B[T]} {\rm meV}$) in order to distinguish LL features in the sound velocity.
%{\bf OK?}

\subsection{Quantum limit}

At strong magnetic fields, $\Pi^R_{V 0}$ is given by Eq.~(\ref{eq:PIVLLL}). 
Importantly, $\Pi_{V 0}({\bf q},0)$ is independent of the angle $\theta$ between the magnetic field and the propagation direction of the sound, but $\Pi_{V0}({\bf q},\omega+i\Gamma)\propto \cos^2\theta$.

\begin{figure}[H]
\centering{}\includegraphics[width=0.45\textwidth]{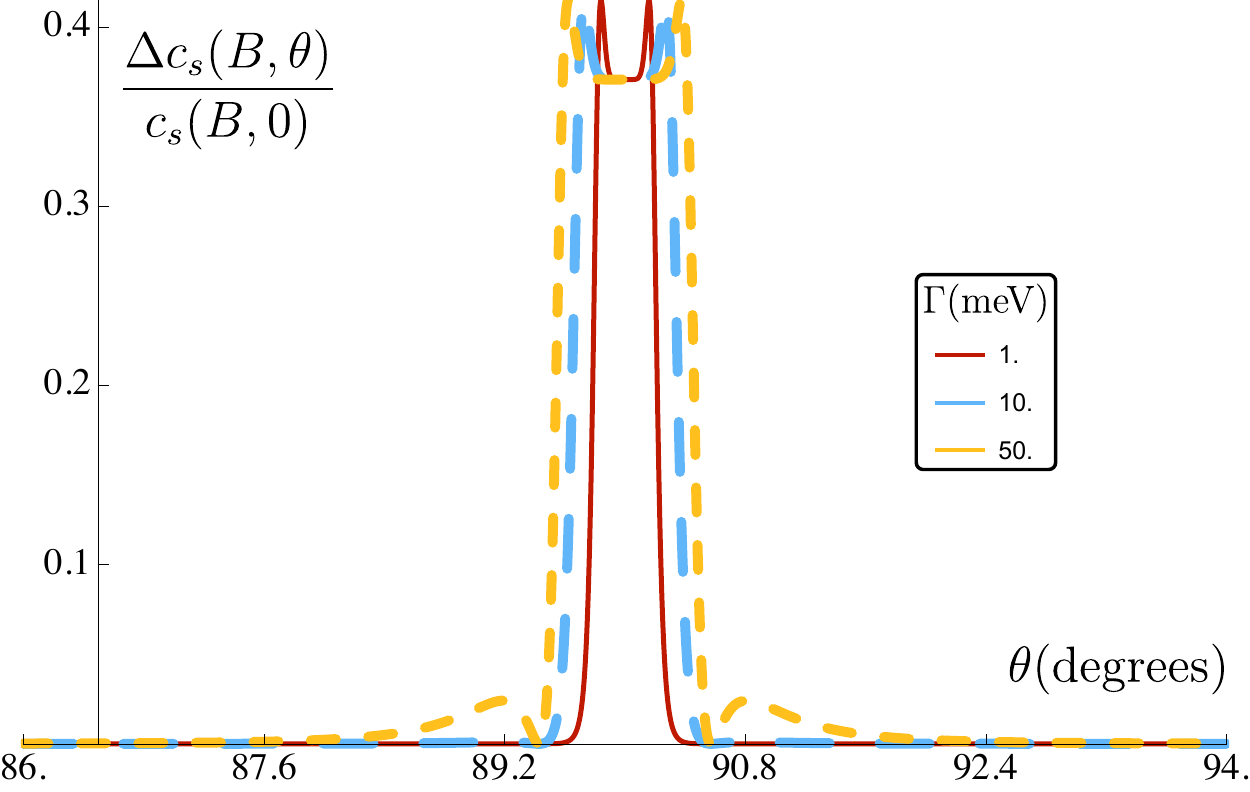}\caption{Dependence of the sound velocity on the relative angle between the magnetic field and the sound propagation direction, in the quantum limit. The vertical axis contains $\Delta c_s(B,\theta) \equiv c_s(B,\theta)-c_s(B,\theta=0)$, where the sound velocity is defined as $|\partial\omega/\partial {\bf q}|$, with $\omega$ given by Eq.~(\ref{eq:dispAC_piezo}). Near $\theta=\pi/2$, the contribution from the chiral LLs is turned off and a peak in the sound velocity ensues. The parameter values are $B=10 {\rm T}$, $\rho=10^4 {\rm kg/m}^3$, $c_0=2 \times 10^3 {\rm m/s}$, $d=3 {\rm C/m}^2$,\cite{Buck}$v_F=10^5 {\rm m/s}$ and $\varepsilon_\infty = 30 \varepsilon_0$.}
\label{fig:cs_gamma}
\end{figure}

Let us assume that $c_s/v_F\gg v_F |{\bf q}|/\Gamma$, which is realistic for the long wavelengths ($|{\bf q}|\simeq 100 {\rm cm}^{-1}$) 
probed in ongoing ultrasound experiments.\cite{Jeffrey}
Consequently, Eq.~(\ref{eq:EQMERMIN}) leads to 
\begin{equation}
\label{eq:epsilon}
%\varepsilon({\bf q},\omega) \simeq 1+\left(\frac{v_F^2 |{\bf q}|^2\cos^2\theta}{\omega^2}-1+i\frac{\Gamma}{\omega}\right)\frac{2\alpha}{\pi} \frac{l^2 \cos^2\theta}{l_B^2},
\varepsilon({\bf q},\omega) \simeq 1+\left(\frac{v_F^2 |{\bf q}|^2 x^2}{\omega^2}-1+i\frac{\Gamma}{\omega}\right)\frac{2\alpha}{\pi} \frac{l^2 x^2}{l_B^2},
\end{equation}
where $l = v_F/\Gamma$ is the electronic mean free path and $x\equiv\cos\theta$.
% and we have used $\omega\simeq c_s|{\bf q}|$.
%It is important to notice the dependence of Eq.~(\ref{eq:epsilon}) on $\cos\theta$.
Keeping everything else fixed, an increase in the magnetic field results in an increase in $|\varepsilon|$ and thus in a reduction of the sound velocity.
Because of the aforementioned link between $\varepsilon$ and $\sigma_{zz}$, the suppresion of the sound velocity in a magnetic field is closely related to the negative magnetoresistance of WSM.
As such, it can be associated to the chiral anomaly, with the advantage of not being mired by current jetting effects.
%It is thus a byproduct of the chiral anomaly.
The $\theta$-dependence of Eq.~(\ref{eq:epsilon}) is likewise nontrivial.
If $\sqrt{\alpha}l |\cos\theta| /l_B \gg {\rm min}\{\sqrt{\omega/\Gamma}, c_s/v_F\}$, then $|\varepsilon|\gg 1$
%imaginary part of the dielectric function is much larger than the real part (which is itself larger than one).
%Therefore, 
and the sound velocity approaches the constant $c_0$.
This corresponds to the situation in which the chiral Landau levels screen the piezoelectric electron-phonon coupling completely.
In contrast, if $|\cos\theta|\lesssim {\rm min} \left\{\frac{l_B}{l \sqrt{\alpha}}\sqrt{\frac{\omega}{\Gamma}}, \sqrt{\frac{c_s}{v_F}\frac{l_B}{l \sqrt{\alpha}}}\right\}$
%if $\sqrt{\alpha}l \cos\theta/l_B \lesssim \sqrt{\omega/\Gamma}$, 
then $\varepsilon({\bf q},\omega)\simeq 1$ and the sound velocity tends to $c_s$ ($c_s>c_0$), i.e. the value it would take in an insulating sample without any chiral LL.
\cite{caveat}
%{\bf and attenuation?} 
Hence, one can switch on and off the chiral Landau level contribution to the sound velocity by varying the angle $\theta$.
In practice, because $\Gamma/\omega$ can be of the order of $10^6$ for $|{\bf q}|\simeq 100{\rm cm}^{-1}$, we are bound to find $c_s\simeq c_0$ for all values of $\theta$ except for a narrow window near $\theta=\pi/2$ (see Fig.~\ref{fig:cs_gamma}).
This window becomes more readily observable in dirtier samples and at higher momenta of the acoustic wave.

\begin{comment}
\[
\sqrt{\left(c^{2}+\text{dc}^{2}\right)}+e^{-\frac{q^{2}\sin^{2}(\theta)}{2B}}\frac{\left(\frac{1}{2}\frac{\text{dc}^{2}Fq\sin^{2}(\theta)}{vq\cos^{2}\theta-i\frac{c_{s}}{v}\Gamma}+\frac{1}{2}\frac{F\left(2(q-2ir\Gamma)c^{2}+\text{dc}^{2}(2q-3ir\Gamma)+2\left(c^{2}+\text{dc}^{2}\right)q\cos(2\theta)\right)B}{q(vq\cos^{2}\theta-i\frac{c_{s}}{v}\Gamma)^{2}}\right)\cos^{2}(\theta)}{\sqrt{\left(c^{2}+\text{dc}^{2}\right)}}
\]
\\
\begin{align*}
 & \left(e^{-\frac{q^{2}\sin^{2}(\theta)}{2B}}\frac{F\cos^{2}(\theta)\left(\frac{\text{dc}^{2}q\sin^{2}(\theta)}{\cos(2\theta)q+q-2ir\Gamma}+\frac{2\left(2(q-2ir\Gamma)c^{2}+\text{dc}^{2}(2q-3ir\Gamma)+2\left(c^{2}+\text{dc}^{2}\right)q\cos(2\theta)\right)B}{q(\cos(2\theta)q+q-2ir\Gamma)^{2}}\right)}{c_{s}}+c_{s}\right)\left(1-2e^{-\frac{q^{2}\sin^{2}(\theta)}{2B}}\frac{F\cos^{2}(\theta)B}{q^{2}\cos^{2}(\theta)-iqr\Gamma}\left[\frac{3+c_{s}^{(0)2}/v^{2}}{4}\right]\right)\\
 & =c_{s}+F\cos^{2}(\theta)\left(e^{-\frac{q^{2}\sin^{2}(\theta)}{2\ell_{B}^{-2}}}\frac{\frac{1}{2}\frac{\text{dc}^{2}vq\sin^{2}(\theta)}{vq\cos^{2}\theta-i\frac{c_{s}}{v}\Gamma}+\frac{\left(2(vq\cos^{2}\theta-ir\Gamma)c_{s}^{2}-ir\Gamma c_{s}^{2}+(2c^{2}vq\cos^{2}\theta-c^{2}ir\Gamma)\right)}{2q(vq\cos^{2}\theta-i\frac{c_{s}}{v}\Gamma)^{2}}\ell_{B}^{-2}}{c_{s}}-2c_{s}e^{-\frac{q^{2}\sin^{2}(\theta)}{2\ell_{B}^{-2}}}\frac{v^{2}\ell_{B}^{-2}}{v^{2}q^{2}\cos^{2}(\theta)-iqc_{s}\Gamma}\left[\frac{3+c_{s}^{(0)2}/v^{2}}{4}\right]\right)
\end{align*}
\end{comment}
{} %
\begin{comment}
$\left(\frac{1}{\rho}\frac{d^{2}}{\varepsilon_{\infty}}/c_{s}^{2}\right)\left(\frac{\alpha}{\pi}\frac{v^{2}\ell_{B}^{-2}}{\left(v\left|\mathbf{q}\right|-i\frac{c_{s}}{v}\Gamma\right)^{2}}\right)+O\left(\left[\frac{v^{2}\ell_{B}^{-2}}{\left(v\left|\mathbf{q}\right|-i\frac{c_{s}}{v}\Gamma\right)^{2}}\right]^{2}\right)$
\end{comment}
%{} and decrease in %
\begin{comment}
$\sqrt{c_{s}^{(0)2}}+\frac{\text{dc}^{2}q\left(3vq-2i\frac{c_{s}}{v}\Gamma/\cos^{2}(\theta)\right)}{2B\sqrt{c^{2}}F}+O\left(\left(\frac{1}{B}\right)^{2}\right)$
\end{comment}
\subsection{Intermediate magnetic fields}

\begin{figure}[H]
%\centering{}\includegraphics[width=0.5\textwidth]{CS_oscil_RTA_BIS2}
\centering{}\includegraphics[width=0.5\textwidth]{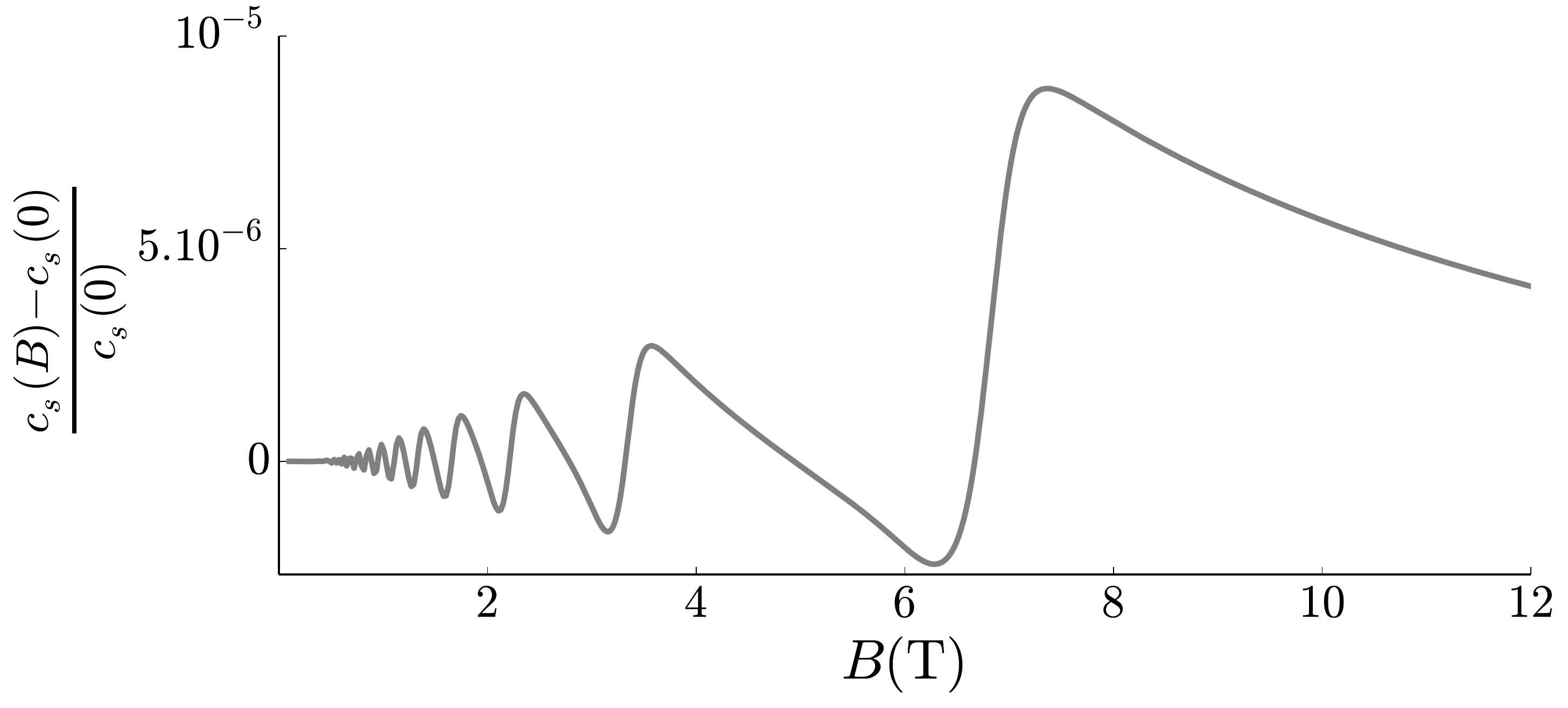}
\caption{Quantum oscillations in the sound velocity, for the case in which the magnetic field is parallel to the sound propagation direction. 
 In the vertical axis, $c_s(B)$ is the sound velocity at magnetic field $B$, while $c_s(0)$ is the sound velocity at zero magnetic field. 
The sound velocity is defined as $|\partial\omega/\partial {\bf q}|$, with $\omega$ given by Eq.~(\ref{eq:dispAC_piezo}).
The zero field value of the sound velocity is taken approximately as $c_s(B=0.1\rm{T})$.
%The envelope of the oscillations scales approximately as $B^{3/2}$, which is expected for Weyl fermions (see Appendix C). 
The parameter values used in the figure read $\theta=0$, $|\mathbf{q}|=1000\rm{cm}^{-1}$, $\Gamma=1 {\rm meV}$, $\rho=10^4\rm{kg/m}^3$, $\varepsilon_\infty=30\varepsilon_0$, $v_F=10^5\rm{m/s}$, $c_0=2\times 10^3\rm{m/s}$ and $d=5\rm{C/m}^2$, with a temperature of $1 {\rm K}$.
\label{fig:cs_oscil}
}
\end{figure}

Away from the quantum limit, the number of nonchiral LLs at the Fermi energy varies with the magnetic field strength.
This results in quantum oscillations for the sound velocity (Fig. \ref{fig:cs_oscil}), which can be detected in high precision ultrasound measurements.\cite{Jeffrey}
% illustrates quantum oscillations of the sound velocity in the intermediate field regime.
Appendix D shows some technical details concerning the characteristics of these oscillations.
%differ from those of simple metals with parabollic dispersion.
%is beyond the scope of this work.
%At low temperature and in the regime $\omega\ll v_F |{\bf q}|\ll \Gamma < v_F/l_B$, the oscillations are 
%The amplitude of the oscillations depends on microscopic parameters such as $d$, $\varepsilon_\infty$ and $\Gamma$.
%The envelope of the oscillations scales approximately as $B^{3/2}$ with the field, 
%As shown in Appendix D, this power law dependence is characteristic of Weyl fermions.
%and differs from a conventional metal at low temperature.\cite{Neuringer}
\textcolor{green}{}
\begin{comment}
\textcolor{green}{Thus, for magnetic field of the order of the Tesla
or more and a chemical potential larger than $\Delta_{1}\sim3\text{meV}\times\sqrt{B(\text{T})}$,
the intraband term from non chiral LL are largely dominant in (\ref{eq:piVstat2}).}
\end{comment}

\section{Summary and Conclusions}
\label{sec:summ}

We have presented a theory that describes the influence of Weyl fermions in the phonon dispersion of disordered Weyl semimetals placed under magnetic fields of arbitrary strength and direction.
Although in principle our formalism is applicable to real Weyl semimetals with multiple pairs of nodes, in practice the task is cumbersome. 
Therefore, in order to extract some simple principles, we have made the following approximations: (i) the electronic dispersion in the absence of magnetic fields has been assumed to be perfectly linear and isotropic in momentum, without tilts; (ii) we have focused on phonon wavevectors that are small compared to the internodal distance; (iii) we have restricted ourselves to scalar and pseudoscalar phonons.
In spite of these approximations, we expect that the physical phenomena predicted in our work will hold in more realistic settings.

The main result from this work is the prediction of a magnetic-field-induced hybridization between plasmons and optical phonons in undoped or weakly doped Weyl semimetals. 
This hybridization can be attributed to an effective charge that optical phonons acquire when interacting with Weyl fermions.
%in a magnetic field.
At strong magnetic fields and nonzero phonon momenta, the hybridization takes place for both scalar and pseudoscalar phonons in a generic WSM without particular symmetry requirements. 
At zero phonon momentum, the plasmon-phonon hybridization occurs only for pseudoscalar phonons; this effect can be traced to the chiral anomaly. 
At strong magnetic fields, the hybridization gap can attain $0.5 {\rm meV}$, thereby making the effect potentially observable e.g. in high-precision Raman experiments. 
Strong deformation potentials, small Fermi velocities, and the presence of numerous pairs of degenerate Weyl nodes may make the hybridization gap larger.  

The hybridization between plasmons and long-wavelength acoustic phonons is practically inexistent because plasmons propagate much faster than the sound.
Nevertheless, we find that Weyl fermions can leave distinctive fingerprints in the sound velocity. 
For example, at strong magnetic fields, the contribution from the chiral Landau levels to the sound velocity can be turned on and off by varying the relative angle between the sound propagation and the magnetic field. 
For the case of piezoelectric electron-phonon interaction, this implies a higher sound velocity when the magnetic field is perpendicular to the magnetic field, because the system is less conducting in that case.
Similarly, for a given direction of the magnetic field, increasing its magnitude results in a reduction of the sound velocity. 
This effect can be regarded as a manifestation of the chiral anomaly.
At weaker fields, we find quantum oscillations in the sound velocity.
%Under certain conditions, the amplitude of the oscillations scales as $B^{3/2}$ with the magnetic field, this power law being characteristic of Weyl fermions.

The present work complements earlier studies conducted by us \cite{Us} and others,\cite{Song} where the influence of the chiral anomaly in the phonon dispersion was first investigated.  
In Ref.~[\onlinecite{Us}], the magnetic field was considered only in linear response and the effects of the chemical potential, temperature and disorder were neglected. 
In the appropriate limits, our current results reduce to those of Ref.~[\onlinecite{Us}].
However, the present work goes beyond Ref.~[\onlinecite{Us}] by treating the magnetic field non perturbatively and by including the effects of finite chemical potential and disorder.
An added bonus from the present approach is the elucidation that the bosonic excitation that hybridizes with the optical phonon evolves from a pseudoscalar boson (i.e. a pole of $\Pi_A$) at low magnetic fields, to a plasmon (i.e. a zero of the dielectric function)  at high magnetic fields. 
In addition, the present work serves to straighten two errors committed in Ref.~[\onlinecite{Us}]: first, phonons of all symmetries (not just $A_1$) are able to scatter electrons in the vicinity of a Weyl node; second, the pseudoscalar electron-phonon can arise in all crystal classes (not just in those where all mirror symmetries are broken).

Concerning Ref. [\onlinecite{Song}], it must be noted that a hybridization between plasmons and optical phonons was clearly predicted therein. 
However, the calculation of Song {\em et al.} was phenomenological and did not evaluate microscopic response functions with the influence from Landau levels.
Perhaps for that reason, the hybridization gap obtained in this work differs from that of Ref. [\onlinecite{Song}].
Besides, Ref. [\onlinecite{Song}] did not recognize that the plasmon-phonon hybridization at nonzero momenta and strong magnetic fields is not unique to pseudoscalar phonons; our theory shows that it occurs for scalar phonons as well.
In fact, for scalar phonons, our theory allows for a hybridization with plasmons at finite momenta even at zero field.
The underlying reason for this is that the response function $\Pi_V$ is nonzero at vanishing magnetic fields and contributes to the effective phonon charge at nonzero momenta.\cite{Us}
What is more unique to WSM is the phonon hybridization gap at {\em zero} momentum, which occurs only for pseudoscalar phonons and only at nonzero magnetic fields. 
The underlying reason for this is that the response function $\Pi_A$ vanishes at zero magnetic field, and generates the chiral anomaly at first order in the magnetic field. 

A possible critique to our work is that some of its results are not exclusive to Weyl semimetals.
For example, group theory predicts that magnetic-field-induced phonon effective charges can arise in many crystals (irrespective of their band topology).
Hence, we cannot rule out (though we are not aware of) a magnetically induced phonon-plasmon hybridization in an intrinsically nonpolar semiconductor or semimetal not hosting any Weyl fermion.

It is our hope that the present work will stimulate further theoretical and experimental studies of electron-phonon interactions in Weyl semimetals. 
On the theory side, it would be particularly valuable to evaluate the plasmon-phonon hybridization gap from first principles.
A first step towards this goal would be to carry out an {\em ab initio} computation of scalar and pseudoscalar deformation potentials in real Weyl semimetals.
In parallel, it would be desirable to conduct experiments that might explore the nontrivial interplay between Landau levels and phonons in Weyl semimetals. 
It would also be useful to complement our calculations with semiclassical methods.

\acknowledgements
This research has been financed by the Canada First Research Excellence Fund, the Natural Science and Engineering Council of Canada, and the Fonds de Recherche du Qu\'ebec Nature et Technologies. We are grateful to S. Acheche, S. Bertrand, J. Quilliam and X. Yuan for helpful conversations.

\appendix

\begin{widetext}

\section{Derivation of Eq.~(\ref{eq:GenDisp})}

In this appendix, we derive the equation governing the phonon dispersion in Weyl semimetals under a magnetic field. 
This equation of motion can be obtained via the minimization of an effective action for phonons with respect to the phonon normal coordinate $v$. 
To simplify the calculation, we will limit the discussion to a single phonon mode.

The imaginary time effective action for phonons, $S_{\rm eff}[v]$, is defined through
\begin{equation}
\label{eq:effac}
e^{-S_{\rm eff}[v]} = e^{-S_0[v]} \int {\rm D} \psi {\rm D} \bar{\psi} \exp\left[-\int d^4 x \bar{\psi} (\partial_t-\mu) \psi -\int dt\left({\cal H}^{(0)}_{\rm el}[\psi,\bar{\psi}] + {\cal H}_{\rm el-el}[\psi,\bar{\psi}]+{\cal H}_{\rm el-ph}[\psi,\bar{\psi},v]\right)\right],
\end{equation}
where $x=({\bf r},t)$, $t$ is the imaginary time, $\mu$ is the chemical potential,
\begin{equation}
\label{eq:bareph}
S_0[v] = \frac{M}{2}  \sum_{\bf q} T\sum_{\omega_n}(\omega_n^2+\omega_{\bf q}^2) v_{\bf q}(\omega_n) v_{-{\bf q}}(-\omega_n) 
\equiv \frac{M}{2}T\sum_{q} (\omega_n^2+\omega_{\bf q}^2) v(q) v(-q)
\end{equation}
is the action for phonons in the absence of Weyl fermions, $\omega_n$ is a Matsubara frequency, $M$ is the atomic mass in a unit cell, $q=({\bf q},\omega_n)$,  $\psi$ and $\bar{\psi}$ are the fermionic Grassmann fields, ${\cal H}_{\rm el}^{(0)}$ is the Hamiltonian of free Weyl fermions in a magnetic field, and ${\cal H}_{\rm el-el}$ (${\cal H}_{\rm el-ph}$) is the electron-electron (electron-phonon) interaction Hamiltonian. 
%In Eq.~(\ref{eq:bareph}), 
The Born effective charge in the absence of Weyl fermions has been assumed to be zero. 
We will relax this assumption at the end of this appendix.
%absorbed in the bare phonon frequency $\omega_{\bf q}$.

\subsection{Free electronic Hamiltonian}

The non-interacting electronic Hamiltonian in the absence of magnetic fields
% ${\bf B}$ 
can be generally written as
\begin{equation}
h=\frac{{\bf p}^2}{2 m} + U({\bf r}) +\frac{\hbar}{4 m^2 c^2} (\boldsymbol{\nabla}U \times{\bf p})\cdot{\bf s},% +\mu_B {\bf s}\cdot{\bf B},
\end{equation}
where 
%${\bf A}$ is the vector potential,  
$U({\bf r})$ is the periodic lattice potential, $m$ is the bare electron mass, $c$ is the speed of light in vacuum, 
%$\mu_B$ is the Bohr magneton, 
${\bf s}$ is a vector of Pauli matrices for the (true) spin, and ${\bf p}=-i\hbar\boldsymbol{\nabla}$ is the momentum operator. 

Weyl semimetals contain pairs of bands that cross at Weyl nodes labeled by an index $\tau$ and located at momenta ${\bf k}_\tau$. 
The low-energy effective model is obtained by expanding the electronic field operators around Weyl nodes as
\begin{equation}
\label{eq:ferfieldexp}
\hat{\psi}^{\dagger}(\mathbf{r}) \simeq\frac{1}{\sqrt{\cal V}}\sum_{\sigma\tau}\text{e}^{i\mathbf{k}_{\tau}\cdot\mathbf{r}}\sum_{\left|\mathbf{k}\right|<\Lambda} e^{i\mathbf{k}\cdot\mathbf{r}}u_{\sigma\tau}^{*}(\mathbf{r})c_{\mathbf{k}\sigma\tau}^{\dagger}
\equiv\sum_{\sigma\tau} e^{i\mathbf{k}_{\tau}\cdot\mathbf{r}} u_{\sigma\tau}^{*}(\mathbf{r}) \hat{\Psi}_{\sigma\tau}^{\dagger}({\bf r}),
\end{equation}
where ${\cal V}$ is the crystal volume, $\Lambda$ is an ultraviolet momentum cutoff,  $\sigma$ labels the two degenerate bands at the Weyl nodes, $u_{\sigma\tau}({\bf r})$ is the periodic part of the Bloch state at momentum ${\bf k}_\tau$ and band $\sigma$,  $c^\dagger_{{\bf k}\sigma\tau}$ creates a Bloch state $\exp(i ({\bf k}_\tau+{\bf k})\cdot{\bf r}) u_{\sigma\tau}({\bf r})$  and  
\begin{equation}
\label{eq:Psi}
\hat{\Psi}_{\sigma\tau}^{\dagger}(\mathbf{r})=\frac{1}{\sqrt{\cal V}}\sum_{\left|\mathbf{k}\right|<\Lambda}c_{\mathbf{k}\sigma\tau}^{\dagger}\text{e}^{i\mathbf{k}\cdot\mathbf{r}}.
\end{equation}
In Eq.~(\ref{eq:ferfieldexp}), $u^*_{\sigma\tau}({\bf r}) \equiv (\langle \uparrow, {\bf r} | u_{\sigma\tau}\rangle^*, \langle \downarrow, {\bf r} | u_{\sigma\tau}\rangle^*)$ is a two-component spinor, where $\uparrow$ and $\downarrow$ denote the real spin projections.

A low-energy effective Hamiltonian for the Weyl nodes can be evaluated by taking the matrix elements of $h$ on the states $\exp(i({\bf k}_\tau+{\bf k})\cdot{\bf r}) u_{\sigma\tau}({\bf r})$. 
In the simplest (two-node) case and to lowest order in ${\bf k}$, this yields
\begin{align}
\label{eq:hel}
{\cal H}^{(0)}_{\rm el} \simeq&\sum_{\sigma\sigma'\tau} \tau \int\text{d}^{3}r {\hat{\Psi}}_{\sigma\tau}^{\dagger}(\mathbf{r})\left[b_0 \delta_{\sigma\sigma'}+\hbar v_F\boldsymbol{\sigma}_{\sigma\sigma'}\cdot\mathbf{k}\right] {\hat{\Psi}}_{\sigma'\tau}(\mathbf{r})\nonumber\\
&\equiv \int\text{d}^{3}r {\hat{\Psi}}^{\dagger}(\mathbf{r})\left[b_0 \tau_z+\hbar v_F\boldsymbol{\sigma}\cdot{\bf k}\tau_z \right] {\hat{\Psi}}(\mathbf{r}),
\end{align}
where $v_F$ is the slope of the linear Weyl dispersion, $\hat{\Psi}({\bf r})$ is a four-spinor of components $\hat{\Psi}_{\sigma\tau}({\bf r})$, and $\sigma_i$ ($\tau_i$) are Pauli matrices in the pseudospin (node) space. 
In the presence of a magnetic field, we neglect the Zeeman splitting and simply replace $\hbar {\bf k}$ by ${\bf p}-e {\bf A}$ in Eq.~(\ref{eq:hel}), where ${\bf A}$ is the vector potential and $e$ is the charge of the electron.

\subsection{Electron-phonon interaction \label{subsec:Electron-phonon-interaction}}

As derived in Ref.~[\onlinecite{Us}], the deformation potential interaction between long-wavelength phonons and Weyl fermions can be written as
% (see Eq. S.31 in the supplemental material):
\begin{equation}
{\cal H}_{\rm el-ph}\simeq\sum_{\mathbf{k}\mathbf{q}}\sum_{\sigma\tau\sigma'} g_{\sigma\sigma'\tau}({\bf q})v_{\mathbf{q}}(t) c_{\mathbf{k}\sigma\tau}^{\dagger}c_{\mathbf{k+q}\sigma\tau'},
\label{eq:Hep1}
\end{equation}
where the momentum sums are constrained within the ultraviolet cutoff, 
\begin{equation}
g_{\sigma\sigma^{'},\tau}({\bf q})=\frac{1}{\sqrt{N}\mathcal{V}_{\rm cell}}\sum_{s}\int_{\rm cell}\text{d}^{3}r\,u_{\sigma\tau}^{*}(\mathbf{r})u_{\sigma\tau}(\mathbf{r})\mathbf{p}_{\mathbf{q}\lambda s}\cdot\frac{\partial U(\mathbf{r}-\mathbf{t}_{s})}{\partial\mathbf{t}_{s}}\text{e}^{-i\mathbf{q\cdot r}},\label{eq:gq1}
\end{equation}
${\cal V}_{\rm cell}$ is the unit cell volume, $s$ is the atom index within the unit cell, ${\bf p}_{{\bf q}s}$ is the polarization vector of atom $s$ and ${\bf q}$ is the phonon momentum. 
%{\bf add factor of mass?}
Here, 
$u_{\sigma\tau}^{*}(\mathbf{r})u_{\sigma^{'}\tau}(\mathbf{r})=\langle u_{\sigma\tau} | {\bf r},\uparrow\rangle \langle {\bf r},\uparrow| u_{\sigma'\tau}\rangle + \langle u_{\sigma\tau} | {\bf r},\downarrow\rangle \langle {\bf r},\downarrow| u_{\sigma'\tau}\rangle$.
%Equation (\ref{eq:gq1}), which was derived in the absence of magnetic fields, remains applicable in the presence of magnetic fields provided that the following two conditions are satisfied: (1) the inverse of the magnetic length should not exceed the distance between the Weyl nodes, so that the Weyl nodes position or chiralities be still good quantum numbers; (2) the electron-phonon interaction has to be local in space, so that the Peierls phases are cancelled. 
%{\bf need to be careful here}
Inverting Eq.~(\ref{eq:Psi}), we can rewrite Eq.~(\ref{eq:Hep1}) as
\begin{align}
{\cal H}_{\rm el-ph} & =\sum_{\sigma\sigma'\tau}\int\text{d}^{3}\mathbf{r}\int\text{d}^{3}\mathbf{r}'\hat{\Psi}_{\sigma\tau}^{\dagger}(\mathbf{r})\sum_{\mathbf{k}\mathbf{q}}\text{e}^{-i\mathbf{k}\cdot\mathbf{r}}\text{e}^{i\left(\mathbf{k}+\mathbf{q}\right)\cdot\mathbf{r}'}g_{\sigma\sigma',\tau}({\bf q})v_{\mathbf{q}}(t)\hat{\Psi}_{\sigma\tau'}(\mathbf{r}')\label{eq:Hep2}\\
 & =\sum_{\sigma\sigma'\tau}\int\text{d}^{3}\mathbf{r}\hat{\Psi}_{\sigma\tau}^{\dagger}(\mathbf{r})\left(\sum_{\mathbf{q}}\text{e}^{i\mathbf{q}\cdot\mathbf{r}}g_{\sigma\sigma',\tau}({\bf q})v_{\mathbf{q}}(t)\right)\hat{\Psi}_{\sigma\tau'}(\mathbf{r}).
\end{align}
Hereafter, we consider only scalar and pseudoscalar electron-phonon couplings, i.e. $g_{\sigma\sigma',\tau}({\bf q})=\delta_{\sigma\sigma'}g_{\tau}({\bf q})$.
Therefore, 
\begin{equation}
\label{eq:elph}
{\cal H}_{\rm el-ph} 
=\int\text{d}^{3}\mathbf{r}\hat{\Psi}^{\dagger}(\mathbf{r}) \left[\phi_{0}(\mathbf{r},t)+\tau_z \phi_{z}(\mathbf{r},t)\right]\hat{\Psi}(\mathbf{r}),
\end{equation}
where we have omitted identity matrices and we have introduced
\begin{align}
\phi_{0}(\mathbf{r},t) & =\sum_{\mathbf{q}} e^{i\mathbf{q}\cdot\mathbf{r}} \left(\sum_\tau g_{\tau}({\bf q})/2\right)v_{\mathbf{q}}(t) \equiv \sum_{\mathbf{q}}\text{e}^{i\mathbf{q}\cdot\mathbf{r}} g_0({\bf q}) v_{\mathbf{q}}(t) = \frac{1}{\mathcal{V}}\sum_{\mathbf{q}}\text{e}^{i\mathbf{q}\cdot\mathbf{r}} \phi_0({\bf q},t)\label{eq:phi01}\\ 
\phi_{z}(\mathbf{r},t) & =\sum_{\mathbf{q}} e^{i\mathbf{q}\cdot\mathbf{r}}\left(\sum_\tau \tau g_{\tau}({\bf q})/2\right)v_{\mathbf{q}}(t) 
\equiv \sum_{\mathbf{q}} e^{i\mathbf{q}\cdot\mathbf{r}} g_z({\bf q}) v_{\mathbf{q}}(t)=\frac{1}{\mathcal{V}}\sum_{\mathbf{q}} e^{i\mathbf{q}\cdot\mathbf{r}} \phi_z({\bf q},t)\label{eq:phiz1}
\end{align}
as the scalar and pseudoscalar deformation potentials, respectively.

At this point, it is useful to discuss symmetry properties of the electron-phonon coupling.
The first question we address is this: what kind of long-wavelength phonons are susceptible to couple to Weyl fermions?
The answer is {\em all phonons}, irrespective of their symmetry.
This answer corrects an error made in Ref.~[\onlinecite{Us}], where we argued that only $A_1$ phonons could scatter Weyl fermions in the vicinity of the Weyl node. 
%This claim is not correct.
%The correct statement would be that {\em all} phonons are able to scatter Weyl fermions in the vicinity of a Weyl node.
Let us elaborate. 
Consider a phonon mode $\lambda$ of momentum ${\bf q}$, which scatters an electron from momentum ${\bf k}_i$ to momentum ${\bf k}_f = {\bf k}_i +{\bf q}$.
According to group theory,\cite{Bassani, Batanouny} the scattering matrix element is generically nonzero provided that
\begin{equation}
\label{eq:borysenko}
\sum_{g\in G_{{\bf k}_i} \cap G_{\bf q} \cap G_{{\bf k}_f}} \left[\chi_{{\bf k}_i}^{(i)} (g)\right]^* \chi^\lambda_{\bf q} (g) \chi^{(f)}_{{\bf k}_f} (g)\neq 0.
\end{equation} 
Here, $\chi_{{\bf k}_i}^{(i)}$ and $\chi_{{\bf k}_f}^{(f)}$ are the characters for the irreducible representations according to which the electronic states at ${\bf k}_i$ and ${\bf k}_f$ transform. 
Also, $\chi_{\bf q}^\lambda$ is the character for the irreducible representation according to which the phonon transforms. 
Importantly, the sum in Eq.~(\ref{eq:borysenko}) contains only those elements $g$ of the crystalline point group which belong {\em simultaneously} to the little groups at ${\bf k}_i$, ${\bf k}_f$ and ${\bf q}$.
These little groups are denoted as $G_{{\bf k}_i}$, $G_{{\bf k}_f}$ and $G_{\bf q}$, respectively.  
For a Weyl node located at an arbitrary position at the Brillouin zone, the only symmetry element in $G_{{\bf k}_i}$ and $G_{{\bf k}_f}$ is the identity $E$.
Hence, the sum in Eq.~(\ref{eq:borysenko}) contains a single term.  
This term is nonzero regardless of the type of phonon, because $\chi^\lambda_{\bf q} (E)$ is nonzero for all $\lambda$ (and equal to the dimension of the irreducible reperesentation according to which the phonon transforms). 

Another symmetry-related question is the following: what kind of long-wavelength phonons can lead to nonzero $g_0$ or $g_z$?
In order to answer this question, let us investigate how $g_\tau$ transforms under a symmetry operation of the crystal (which we will denote $\hat{R}$).
We will beging by writing the expression for $g_{\hat{R}\tau}$, and attempt to relate it to $g_\tau$ through the following manipulations:
\begin{align}
g_{\hat{R} \tau}({\bf q}) &= \frac{1}{\sqrt{N}\mathcal{V}_{\rm cell}}\int d^3 r e^{i {\bf q}\cdot{\bf r}} \sum_\sigma |u_{\sigma, \hat{R}\tau}({\bf r})|^2 \sum_s {\bf p}_{{\bf q}\lambda s}\cdot\frac{\partial U({\bf r}-{\bf t}_s)}{\partial {\bf t}_s}\\
&=\frac{1}{\sqrt{N}\mathcal{V}_{\rm cell}}\int d^3 r e^{i{\bf q}\cdot{\bf r}} \sum_\sigma |u_{\sigma\tau}(\hat{R}^{-1} {\bf r})|^2 \sum_s {\bf p}_{{\bf q}\lambda s}\cdot\frac{\partial U({\bf r}-{\bf t}_s)}{\partial {\bf t}_s}\label{eq:1}\\
&=\frac{1}{\sqrt{N}\mathcal{V}_{\rm cell}}\int d^3 r e^{i {\bf q}\cdot \hat{R}{\bf r}} \sum_\sigma |u_{\sigma\tau}({\bf r})|^2 \sum_s {\bf p}_{{\bf q}\lambda s}\cdot\frac{\partial U(\hat{R}{\bf r}-{\bf t}_s)}{\partial {\bf t}_s}\label{eq:2}\\
&=\frac{1}{\sqrt{N}\mathcal{V}_{\rm cell}}\int d^3 r e^{i {\bf q}\cdot \hat{R}{\bf r}} \sum_\sigma |u_{\sigma\tau}({\bf r})|^2 \sum_s {\bf p}_{{\bf q}\lambda \hat{R}s}\cdot\frac{\partial U(\hat{R}{\bf r}-\hat{R}{\bf t}_s)}{\partial \hat{R}{\bf t}_s}\label{eq:3}\\
&=\frac{1}{\sqrt{N}\mathcal{V}_{\rm cell}}\int d^3 r e^{i \hat{R}^{-1} {\bf q}\cdot{\bf r}} \sum_\sigma |u_{\sigma\tau}({\bf r})|^2 \sum_s {\bf p}_{{\bf q}\lambda \hat{R}s}\cdot\hat{R}^{-1}\frac{\partial U({\bf r}-{\bf t}_s)}{\partial {\bf t}_s}\label{eq:4}\\
&=\frac{1}{\sqrt{N}\mathcal{V}_{\rm cell}}\int d^3 r e^{i \hat{R}^{-1} {\bf q}\cdot{\bf r}} \sum_\sigma |u_{\sigma\tau}({\bf r})|^2 \sum_s \left(\hat{R}\,{\bf p}_{{\bf q}\lambda \hat{R}s}\right)\cdot\frac{\partial U({\bf r}-{\bf t}_s)}{\partial {\bf t}_s}\label{eq:5}.
\end{align}
In Eq.~(\ref{eq:1}), we have used the relation 
%$u_{\sigma \hat{R} {\bf k}}({\bf r}) = \sum_{\sigma'} M_{\sigma\sigma'} u_{\sigma'\bf k}(\hat{R}^{-1} {\bf r})$, where $M$ is a $2\times 2$ unitary matrix.
%It follows that 
$\sum_\sigma |u_{\sigma \hat{R} {\bf k}}({\bf r})|^2 = \sum_\sigma |u_{\sigma {\bf k}}(\hat{R}^{-1} {\bf r})|^2$ (for a justification, see e.g. Ref. [\onlinecite{Bradley}]).
In Eqs.~(\ref{eq:2}) and (\ref{eq:3}), we have simply redefined the dummy variables ${\bf r}$ and $s$.
In particular, a sum over ${\bf t}_s$ equals a sum over $\hat{R} {\bf t}_s$, because $\hat{R}$ is a symmetry operation of the crystal.
In Eq.~(\ref{eq:4}), we have used $U({\bf r}-{\bf t}_s) = U(|{\bf r}-{\bf t}_s|)$, ${\bf q}\cdot \hat{R} {\bf r} = (\hat{R}^{-1} {\bf q})\cdot{\bf r}$ and $\hat{R}^{-1} \partial_{{\bf t}_s} = \partial_{\hat{R}{\bf t}_s}$.
%In Eq.~(\ref{eq:5}), we have invoked the fact that that a gradient transforms as a vector under symmetry operations.
Then, we conclude
\begin{equation}
g_{\hat{R}\tau}(\hat{R}{\bf q}) =\frac{1}{\sqrt{N}\mathcal{V}_{\rm cell}}\int d^3 r e^{i{\bf q}\cdot{\bf r}} \sum_\sigma |u_{\sigma\tau}({\bf r})|^2 \sum_s \left(\hat{R}\,{\bf p}_{(\hat{R}{\bf q})\lambda \hat{R}s}\right)\cdot\frac{\partial U({\bf r}-{\bf t}_s)}{\partial {\bf t}_s}.
\end{equation}
Let us now take $\hat{R}$ to be a mirror operation.
In the small momentum regime (${\bf q}\simeq \hat{R}{\bf q}\simeq 0$), a phonon is either odd or even under a mirror transformation: $\hat{R}\,{\bf p}_{{\bf 0}\lambda \hat{R}s} \simeq (-1)^\lambda{\bf p}_{{\bf 0}\lambda s}$, where $\lambda=+(-) 1$ for a scalar (pseudoscalar) mode.
Thus, $g_{-\tau}({\bf q}) \simeq (-1)^\lambda g_\tau ({\bf q})$ in the $q\to 0$ limit (this can hold at nonzero ${\bf q}$ as well, provided that the phonon momentum is located at a $\hat{R}$-invariant plane).
Consequently, long-wavelength scalar (pseudoscalar) phonons couple to Weyl fermions predominantly via $g_0\propto\sum_\tau g_\tau$ ($g_z\propto\sum_\tau \tau g_\tau$).

In Ref.~[\onlinecite{Us}], it was argued that $g_z\neq 0$ requires that all mirror symmetries be broken.
This statement was based on the assumption that $\sum_\sigma|u_{\sigma\tau}({\bf r})|^2 = \sum_\sigma |u_{\sigma, \hat{R}\tau} ({\bf r})|^2$, or equivalently $\sum_\sigma|u_{\sigma\tau}({\bf r})|^2 = \sum_\sigma |u_{\sigma\tau} (\hat{R}^{-1}{\bf r})|^2$.
However, this assumption is generally incorrect because the probability density of a particular Bloch state (with fixed but otherwise arbitrary crystal momentum) need not obey the symmetries of the Hamiltonian.
%{\bf say more here?}

\subsection{Electron-electron interaction}

The electron-electron interactions are described by 
\begin{equation}
{\cal H}_{\rm el-el}=\frac{1}{2} \int d^3 r d^3 r' V({\bf r}-{\bf r}') \hat{\psi}^\dagger({\bf r}) \hat{\psi}^\dagger({\bf r}') \hat{\psi}({\bf r}') \hat{\psi}({\bf r}),
\end{equation}
where $V({\bf r}) = e^2/(4 \pi \varepsilon_0 r)$ is the Coulomb potential.
In the low-energy effective theory, we expand $\hat{\psi}$ in the truncated Hilbert space of the two bands that cross at the Weyl nodes, and replace $\varepsilon_0$ by $\varepsilon_\infty$ in order to incorporate the screening effects coming from bands outside the truncated Hilbert space. 
Then, keeping only the long-wavelength part of the Coulomb interaction and neglecting umklapp terms, we arrive at  
\begin{equation}
\label{eq:elel}
{\cal H}_{\rm el-el}\simeq\frac{1}{2}\frac{1}{\mathcal{V}}\sum_{\mathbf{q}} V({\bf q})\rho\left(\mathbf{q}\right)\rho\left(-\mathbf{q}\right),
\end{equation}
where $V({\bf q}) = e^2/(\varepsilon_\infty {\bf q}^2)$ and $\rho\left(\mathbf{q}\right)\equiv\sum_{\left|\mathbf{k}\right|<\Lambda}\sum_{\sigma\tau} c_{\mathbf{k}\sigma\tau}^{\dagger}c_{\mathbf{k}+\mathbf{q}\sigma\tau}$ is the low-energy electronic density operator.

\subsection{Integrating out electrons}

To get an effective action for phonons, we need to integrate over the fermionic fields in the right hand side of Eq.~(\ref{eq:effac}).
In order to treat the quartic term in fermionic fields coming from ${\cal H}_{\rm el-el}$,  we use the Hubbard-Stratonovich transformation 
\begin{equation}
\exp\left[{-\frac{1}{2}\frac{T}{\mathcal{V}}\sum_q V({\mathbf{q}}) \rho(q) \rho(-q)}\right]
=
\frac{1}{\mathcal{N}}\int\text{D}\varphi\exp\left\{-\frac{T}{\mathcal{V}}\sum_q\left[\frac{\varepsilon_{\infty}\mathbf{q}^{2}}{2}\varphi\left(q\right)\varphi\left(-q\right)+ie\varphi\left(q\right)\rho\left(-q\right)\right]\right\},
\label{eq:Hub-Strat1}
\end{equation}
where $\mathcal{N}$ is a constant, T is the temperature in unit of frequency and the auxiliary bosonic field $i \varphi$ can be interpreted as the electric potential created by the fluctuations of the electronic density (such that the minimization of the exponent in Eq. (\ref{eq:Hub-Strat1}) with respect to $\varphi$ gives the Poisson equation).

Combining Eqs.~(\ref{eq:hel}), (\ref{eq:elph}), (\ref{eq:elel}) and (\ref{eq:Hub-Strat1}),  Eq.~(\ref{eq:effac}) can be written as
\begin{equation}
e^{-S_{\rm eff}[v]} \simeq \frac{1}{\cal N} e^{-S_0[v]} \int D\varphi  D\Psi D\bar{\Psi} 
\exp\left[-\frac{T}{\mathcal{V}}\sum_q \frac{\varepsilon_{\infty}\mathbf{q}^{2}}{2}\varphi\left(q\right)\varphi\left(-q\right) +\int d^4x \bar{\Psi}(x) G^{-1}(x,x) \Psi(x)\right],
\end{equation}
where 
%$x=({\bf r},t)$ and 
\begin{equation}
G^{-1}(x,x)= G_0^{-1}(x,x) - V(x) 
\end{equation}
is the inverse of the full fermionic Green's function, and the Green's function for Weyl fermions in the absence of interactions is defined as $\hbar\left[-\partial_t-\mu - v_F  \left(-i\boldsymbol{\nabla}-e {\bf A}(x)\right)\cdot{\boldsymbol\sigma}\tau_z - b_0 \tau_z\right]G_0(x,x')=\delta(x-x')$. Additionally,
\begin{equation}
V(x)=i e \varphi(x) + \phi_0(x) + \phi_z(x) \tau_z
\end{equation}
 is the correction due to electron-phonon and electron-electron interactions.
%{\bf Arguments OK for the inverse Green's function?}

The integral over the fermionic fields is now gaussian and can be done exactly.
Assuming the absence of lattice instabilities and electronic ordered states, we can expand the resulting expression to lowest order in powers of $\phi_0$, $\phi_z$ and $\varphi$.\cite{Nagaosa}
The outcome reads
\begin{align}
\label{eq:effac2}
e^{-S_{\rm eff}[v]} 
&\simeq \frac{1}{\cal N} e^{-S_0[v]} \int D\varphi \,{\rm det}(-G^{-1})  \exp\left[-\int\frac{\text{d}^{4}q}{\left( 2\pi\right)^4}\frac{\varepsilon_{\infty}\mathbf{q}^{2}}{2}\varphi\left(q\right)\varphi\left(-q\right)\right]\nonumber\\
%-\frac{1}{2} \int d^4 x d^4 x' {\rm tr}\left[G_0(x,x') V(x') G_0(x',x) V(x)\right]\right],
&\simeq \frac{1}{\cal N'} e^{-S_0[v]} \int D\varphi  \exp\left[-\int\frac{\text{d}^{4}q}{\left( 2\pi\right)^4}\frac{\varepsilon_{\infty}\mathbf{q}^{2}}{2}\varphi\left(q\right)\varphi\left(-q\right)-\frac{\hbar}{2} \int d^4 x \,d^4 x' {\rm tr}\left[G_0(x,x') V(x') G_0(x',x) V(x)\right]\right],
\end{align}
where ${\cal N}'$ is another constant,  and ${\rm tr}$ stands for trace in the $(\sigma,\tau)$.
%\begin{equation}
%V(x)\equiv \phi_0(x) + i e \varphi(x) + \phi_z(x) \tau_z
%\end{equation}

Following Schwinger,\cite{Schwinger} the Green's function of noninteracting Weyl fermions in a magnetic field can be factorized as
\begin{equation}
G_0(x,x') = {\cal G}_0(x-x') \Phi(x,x'),
\end{equation}
where $\Phi(x,x')$ is a phase factor such that $\Phi(x,x') \Phi(x',x)=1$ and $\mathcal{G}_0(x-x')=\frac{T}{\cal V}\sum_k\mathcal{G}_0(k)e^{ik\cdot x}$.
Then,
\begin{align}
\label{eq:GVGV}
&\hbar\int d^4 x \, d^4 x' {\rm tr}\left[G_0(x,x') V(x') G_0(x',x) V(x)\right] = \hbar\int d^4 x \, d^4 x' {\rm tr}\left[{\cal G}_0(x-x') V(x') {\cal G}_0(x'-x) V(x)\right]\nonumber\\
& = \hbar \frac{T^2}{\mathcal{V}^2}\sum_q \sum_k {\rm tr}\left[{\cal G}_0(k) V(q) {\cal G}_0(k+q) V(-q)\right]\nonumber\\
&=\frac{T}{\mathcal{V}}\sum_q \,\Pi_V(q) \left[(\phi_0(q)+ie \varphi(q))(\phi_0(-q)+ i e \varphi(-q))+\phi_z(q)\phi_z(-q)\right]\nonumber\\
&+\frac{T}{\mathcal{V}}\sum_q \,\Pi_A(q)\left[(\phi_0(q)+ie \varphi(q)) \phi_z(-q) + (\phi_0(-q)+ie \varphi(-q))\phi_z(q)\right],
\end{align}
where
\begin{align}
\Pi_V(q) & \equiv\hbar\frac{T}{\mathcal{V}}\sum_k\text{tr}\left[\mathcal{G}_0(k)\mathcal{G}_0(k+q)\right]=\Pi_V(-q)\label{PiVVapp1}\\
\Pi_A(q) & \equiv\hbar\frac{T}{\mathcal{V}}\sum_k\text{tr}\left[\tau_z\mathcal{G}_0(k)\mathcal{G}_0(k+q)\right]=\Pi_A(-q)
\end{align}
are the scalar and pseudoscalar density response functions.

It is illustrative to reinterpret Eq.~(\ref{eq:GVGV}) by rewriting it as
\begin{align}
&\int d^4 x \, d^4 x' {\rm tr}\left[G_0(x,x') V(x') G_0(x',x) V(x)\right] \nonumber\\
&=\frac{T}{\mathcal{V}}\sum_q \,\Pi_V(q) \left[|\phi_0(q)|^2+|\phi_z(q)|^2\right] + 2 \Pi_A(q){\rm Re}[\phi_0(q) \phi_z(-q)]\nonumber\\
&+\frac{T}{\mathcal{V}}\sum_q {\bf Q}_{\rm el}(q)\cdot{\bf E}(-q) v(q) +{\bf Q}_{\rm el}(-q)\cdot{\bf E}(q) v(-q),
\end{align}
where ${\bf E}(q) = -{\bf q} \varphi(q)$ is a longitudinal electric field and ${\bf Q}_{\rm el}$ is the contribution from itinerant electrons to the effective phonon charge.
The latter  obeys
\begin{equation}
{\bf Q}_{\rm el}(q)\cdot{\bf q} = i\frac{e\cal{V}}{\sqrt{N}}\left[\Pi_V(q) g_0({\bf q}) + \Pi_A(q) g_z({\bf q})\right].
\end{equation}  
Thus, $(i e {\cal V}/\sqrt{N})\Pi_V(q) g_0({\bf q})/|{\bf q}|$ and $(i e {\cal V}/\sqrt{N})\Pi_A(q) g_0({\bf q})/|{\bf q}|$ are the scalar and pseudoscalar contributions to the longitudinal component of the effective phonon charge, respectively.
%This evidences the connection of $\Pi_V$ and $\Pi_A$ with the Born charge.
 
Let us continue with the derivation of the phonon dispersion. 
From Eq.~(\ref{eq:GVGV}), it is evident that the fields $\varphi(q)$ and $v(q)$ are coupled in the right hand side of Eq.~(\ref{eq:effac2}).
We can decouple them by making a change of variables
\begin{equation}
\varphi(q)\rightarrow\varphi(q)-ie\frac{\phi_{0}\left(q\right)\Pi_V(q)+\phi_{z}\left(q\right)\Pi_A(q)}{\varepsilon_\infty\varepsilon(q)\mathbf{q}^2},
\end{equation}
where $\varepsilon\left(q\right)=1-V({\bf q})\text{\ensuremath{\Pi}}_{V}\left(q\right)$ is the contribution from Weyl fermions to the dielectric function.
After some algebra, Eq.~(\ref{eq:effac2}) becomes
\begin{equation} 
e^{-S_{\rm eff}[v]} \simeq \frac{1}{\cal N'} e^{-S_0[v]-\delta S[v]} \int D\varphi \exp\left[-\frac{1}{2}\frac{T}{\mathcal{V}}\sum_q \varepsilon_\infty \varepsilon(q) {\bf q}^2 \varphi(q)\varphi(-q)\right],
\end{equation}
where 
\begin{equation}
\delta S[v] = \frac{1}{2}\frac{T}{\mathcal V}\sum_q
\frac{1}{\varepsilon(q)}\left\{  (|\phi_0(q)|^2+|\phi_z(q)|^2)\Pi_V(q)+2 {\rm Re}(\phi_0(q) \phi_z(-q))\Pi_A(q)-V({\bf q})\left[\Pi_V(q)^{2}-\Pi_A(q)^{2}\right]\left|\phi_{z}(q)\right|^{2}\right\}. 
\end{equation}
Thus, the phonon effective action is 
\begin{equation}
S_{\rm eff}[v]=S_0[v] + \delta S[v] + \text{terms independent of $v$}. 
\end{equation}
Using $\phi_{0(z)}(q) = \mathcal{V}g_{0 (z)}({\bf q})v(q)$ and taking $\delta S_{\rm eff}/\delta v_{\bf q}(q_0) = 0$, we obtain the expression for the phonon dispersion
\begin{equation}
q_0^2 +\omega_{{\rm ph},{\bf q}}^{2}+\frac{{\cal V}}{M\varepsilon(q)}\left\{ (|g_0({\bf q})|^2+|g_z({\bf q})|^2)\Pi_V(q)+ 2 {\rm Re}(g_0({\bf q}) g_z(-{\bf q}))\Pi_A(q)-V({\bf q})\left[\Pi_V(q)^{2}-\Pi_A(q)^{2}\right]\left|g_{z}({\bf q})\right|^{2}\right\}=0. 
\end{equation}
Performing an analytic continuation ($i q_0\to \omega + i 0^+$), we recover the result quoted in the main text.
In particular, for a scalar phonon mode ($g_{z}=0$), we recover the standard phonon
dispersion [\onlinecite{Mahan}]
\[
\omega^{2}=\omega_{{\rm ph},{\bf q}}^{2}+\frac{{\cal V}}{M}\frac{\left|g_{0}({\bf q})\right|^{2}}{\varepsilon(q)}\Pi^R_V(q),
\]
whereas the dispersion becomes 
\[
\omega^{2}=\omega_{{\rm ph},{\bf q}}^{2}+\frac{{\cal V}}{M}\left|g_{z}({\bf q})\right|^{2}\left[\Pi^R_V(q)+V(\mathbf{q})\frac{\Pi^R_A(q)^{2}}{\varepsilon(q)}\right]
\]
for a purely pseudoscalar mode ($g_{0}=0$).

\subsection{Phonon dispersion in the presence of a nonzero bare effective charge}

Thus far, we have studied the case in which the phonon effective charge vanishes in the absence of conduction electrons.
Here, we provide the generalization of the preceding results to the case of polar optical phonons with a nonzero ``bare'' effective charge ${\bf Q}_{\bf q}^{(0)}$.
The main change with respect to the derivation shown above consists of replacing Eq.~(\ref{eq:elel}) by \cite{Mahan}
\begin{equation}
\label{eq:hint}
{\cal H}_{\rm int} = \sum_{\bf q} V({\bf q}) \rho_{\rm tot}({\bf q}) \rho_{\rm tot}(-{\bf q}),
\end{equation}
where $\rho_{\rm tot} = \rho + \rho_{\rm ph}$ and
\begin{equation}
\rho_{\rm ph} = \frac{i\sqrt{N}}{e} ({\bf Q}_{\bf q}^{(0)}\cdot{\bf q}) v(q).
\end{equation}
The presence of $i {\bf q}$ in this equation originates from the fact that the phonon charge can be written as a divergence of the electric polarization induced by the lattice vibrations.
In Eq.~(\ref{eq:hint}), the cross-terms between $\rho$ and $\rho_{\rm ph}$ describe the polar coupling between the optical phonon and electrons.
In addition, the same optical phonon couples to electrons via deformation potential coupling, as captured by Eq.~(\ref{eq:elph}).

%{\bf check if you are double counting electron-phonon interactions}
Next, one performs a Hubbard Stratonovich transformation to treat ${\cal H}_{\rm int}$ and performs a perturbative expansion in $\varphi$ and $\phi$ fields to arrive at
\begin{align}
e^{-S_{\rm eff}[v]} \simeq &\frac{1}{\cal N'} e^{-S_0[v]} \int D\varphi \exp\left[-\int\frac{d^4q}{(2\pi)^4} \left(\frac{\varepsilon_\infty {\bf q}^2}{2} \varphi(q) \varphi(-q) + i e \varphi(q) \rho_{\rm ph}(-q)\right)\right.\nonumber\\
&\left.-\frac{1}{2} \int d^4 x\, d^4 x'\, {\rm tr}\left[G_0(x,x') V(x') G_0(x',x) V(x)\right]\right].
\end{align}
As before, we make an appropriate shift in $\varphi$ in order to decouple it from $v$. 
The resulting effective action for phonons reads $S_{\rm eff}[v] = S_0[v]+ \delta S [v] + \text{terms independent of $v$}$, with
\begin{align}
\delta S[v] =& \frac{1}{2}\frac{T}{\mathcal V}\sum_q
\frac{1}{\varepsilon(q)}\left\{  (|\phi_0(q)|^2+|\phi_z(q)|^2)\Pi_V(q)+2 {\rm Re}(\phi_0(q) \phi_z(-q))\Pi_A(q)-V({\bf q})\left[\Pi_V(q)^{2}-\Pi_A(q)^{2}\right]\left|\phi_{z}(q)\right|^{2}\right.\nonumber\\
&\left.+V({\bf q}) |\rho_{\rm ph}(q)|^2 + 2 V({\bf q}) {\rm Re}\left[\rho_{\rm ph}(q)\Pi_V(q) \phi_0(-q) + \rho_{\rm ph}(q) \Pi_A(q) \phi_z(-q)\right]\right\}.
\end{align}
The phonon dispersion can be obtained once again from $\delta S_{\rm eff}/\delta v_{\bf q}(q_0) = 0$.
The effect of the bare phonon effective charge is collected in the second line of $\delta S$.
The first term in the second line is the usual\cite{Mahan} renormalization of the phonon frequency in presence of a bare effective charge, with the factor $1/\varepsilon(q)$ denoting screening from itinerant electrons.
The remainder of the second line describes the interference terms between the bare effective phonon charge and the electronically induced effective phonon charge.

In sum, the new equation for the phonon dispersion reads
\begin{align}
\label{eq:dispborn}
\omega^{2} & =\omega_{{\rm ph},\mathbf{q}}^{2}+\frac{\mathcal{V}}{M\varepsilon\left(q\right)}\left[\Pi_{V}\left(q\right)\left(\left|g_{0}\right|^{2}+\left|g_{z}\right|^{2}\right)+2\Pi_{A}\left(q\right)\text{Re}\left(g_{0}g_{z}^{*}\right)+V(\mathbf{q})\left|g_{z}\right|^{2}\left(\Pi_{A}\left(q\right)^{2}-\Pi_{V}\left(q\right)^{2}\right)\right]\nonumber\\
 & +\frac{V(\mathbf{q})}{M\varepsilon\left(q\right)}\left(\frac{1}{e^2\mathcal{V}_{\rm cell}}\biggl|\mathbf{Q}_{\mathbf{q}}^{(0)}\cdot\mathbf{q}\biggr|^{2}-2\frac{\sqrt{N}}{e}\left(\Pi_{V}\left(q\right)\text{Im}\left[\mathbf{q}\cdot\mathbf{Q}_{\mathbf{q}}^{(0)}g_{0}^{*}\right]+\Pi_{A}\left(q\right)\text{Im}\left[\mathbf{q}\cdot\mathbf{Q}_{\mathbf{q}}^{(0)}g_{z}^{*}\right]\right)\right),
\end{align}
where we have omitted the momentum label from $g_0$ and $g_z$, for the sake of brevity.
When ${\bf Q}_{\bf q}^{(0)}=0$, we recover Eq.~(\ref{eq:GenDisp}).
In passing, we remark that all terms involving ${\bf Q}_{\bf q}^{(0)}$ remain finite at ${\bf q}\to 0$.

\section{Explicit expressions for the response functions}

In the preceding appendix, we have written the phonon dispersion relation in terms of the response functions $\Pi_V$ and $\Pi_A$.
In this appendix, we show the explicit expressions for $\Pi_V$ and $\Pi_A$, which we have used to obtain the numerical results of the main text.

\subsection{Response functions for $b_0=0$}

We start by considering the case $b_0=0$. 
In this case, the explicit expression for Dirac fermions in a uniform and static magnetic field is known [\onlinecite{Shovkovy}] to be
\begin{align}
\mathcal{G}_0(k)= & 2i e^{-k_{\perp}^{2}\ell_{B}^{2}}\sum_{\nu}\sum_n\frac{(-1)^{n}}{\hbar^{2}\left(i\nu\right)^{2}-E_{nk_{z}}^{2}}\hbar\left(i\nu-\tau_{z}\sigma_{z}v_F k_{z}\right)\left[L_{n}(t)\mathcal{P}_{+}-L_{n-1}(t)\mathcal{P}_{-}\right]+2\left(\tau_{z}\boldsymbol{\sigma}_{\perp}\cdot\mathbf{k}_{\perp}\right)L_{n-1}^{1}(t)\label{eq:SchwinGF},
\end{align}
where $\nu$ is a Matsubara frequency, $n\geq 0$ is the (conduction band) Landau level index, $k=(k_z,{\bf k}_\perp,\nu)$, $t=2k_{\perp}^{2}\ell_{B}^{2}$, $\mathcal{P}_{\pm}=(1\pm\sigma_{z})/2$, and $E_{nk_{z}}\equiv\hbar v_F\sqrt{k_{z}^{2}+2n\ell_{B}^{-2}}$. 
Also, $L_{n}^{\alpha}(t)$ are Laguerre polynomials defined as
\[
L_{n}^{\alpha}(t)=\frac{1}{n!}e^{t}t^{-\alpha}\frac{\text{d}^{n}}{\text{d}t^{n}}\left(e^{-t}t^{n+\alpha}\right),
\]
with $L_{n}(t)=L_{n}^{0}(t)$. 
The Laguerre polynomial $L_{n}^{\alpha}\left(x\right)$ is null for $n<0$, by definition. 

Replacing Eq.(\ref{eq:SchwinGF}) in Eq.(\ref{PiVVapp1}), doing the trace in $\sigma,\tau$ space and performing the sum over Matsubara frequencies, we arrive at
\begin{align}
\Pi_V(q) & =-e^{-\frac{q_{\perp}^{2}\ell_{B}^{2}}{2}}\frac{1}{\cal V}\sum_{nm\lambda\lambda'\mathbf{k}}e^{-2\left(\mathbf{k}_{\perp}+\frac{\mathbf{q}_{\perp}}{2}\right)^{2}\ell_{B}^{2}}\left(-1\right)^{n+m}\frac{f(\lambda E_{n k_{z}})-f(\lambda' E_{mk'_{z}})}{\lambda E_{nk_{z}}-\lambda'E_{mk'_{z}}+i\omega}\label{eq:PiVVgen}\\
 & \times\left\{ \left[L_{n}(t)L_{m}(t')+L_{n-1}(t)L_{m-1}(t')\right]\right.\nonumber \\
 & +\left.\lambda\lambda'\hbar^2v_F^2\frac{k_{z}k'_{z}\left[L_{n}(t)L_{m}(t')+L_{n-1}(t)L_{m-1}(t')\right]+8\mathbf{k}_{\perp}\cdot\mathbf{k}'_{\perp}L_{n-1}^{1}(t)L_{m-1}^{1}(t')}{E_{nk_{z}}E_{mk'_{z}}}\right\} \nonumber 
\end{align}
and
\begin{align}
\Pi_A(q) & =e^{-\mathbf{q}_{\perp}^{2}\ell_{B}^{2}/2}\frac{1}{\cal V}\sum_{nm\lambda\lambda'\mathbf{k}}e^{-2\left(\mathbf{k}_{\perp}+\frac{\mathbf{q}_{\perp}}{2}\right)^{2}\ell_{B}^{2}}(-1)^{n+m}\frac{f\left(\lambda E_{nk_{z}}\right)-f\left(\lambda'E_{mk'_{z}}\right)}{\lambda E_{nk_{z}}-\lambda'E_{mk'_{z}}+i\omega}\label{eq:PIVAgen}\\
 & \times \hbar v\left(\lambda\frac{k_{z}}{E_{nk_{z}}}+\lambda'\frac{k'_{z}}{E_{mk'_{z}}}\right)\left(L_{n}(t)L_{m}(t')-L_{n-1}(t)L_{m-1}(t')\right)\nonumber ,
\end{align}
where $\lambda,\lambda'=\pm$, and primed variables correspond to a translation by $\mathbf{q}$ (for instance, $\mathbf{k}'_{\perp}=\mathbf{k}_{\perp}+\mathbf{q}_{\perp}$).
Equations (\ref{eq:PiVVgen}) and (\ref{eq:PIVAgen}) can be simplified by performing the sum over ${\bf k}_\perp$ analytically via the following formulae:
%For the sum (integral) over $\mathbf{k}_{\perp}$, we use the formulae
\begin{align}
\label{eq:formula1}\frac{1}{\left(2\pi\right)^{2}}\int\text{d}^{2}\mathbf{k}_{\perp}e^{-2\left(\mathbf{k}_{\perp}+\frac{\mathbf{q}_{\perp}}{2}\right)^{2}\ell_{B}^{2}}L_{n}\left(2\mathbf{k}_{\perp}^{2}\ell_{B}^{2}\right)L_{m}\left(2(\mathbf{k}'_{\perp})^2\ell_{B}^{2}\right)=\frac{\ell_{B}^{-2}}{8\pi}\left(-1\right)^{n+m}\frac{m!}{n!}\left(\frac{\mathbf{q}_{\perp}^{2}\ell_{B}^{2}}{2}\right)^{n-m}L_{m}^{n-m}\left(\frac{\mathbf{q}_{\perp}^{2}\ell_{B}^{2}}{2}\right)^{2}
\end{align}
and
\begin{align}\label{eq:formula2}
 & \frac{1}{\left(2\pi\right)^{2}}\int\text{d}^{2}\mathbf{k}_{\perp}e^{-2\left(\mathbf{k}_{\perp}+\frac{\mathbf{q}_{\perp}}{2}\right)^{2}\ell_{B}^{2}}\,8\mathbf{k}_{\perp}\!\cdot\mathbf{k}'_{\perp} L_{n-1}^{1}\left(2\mathbf{k}_{\perp}^{2}\ell_{B}^{2}\right)L_{m-1}^{1}\left(2(\mathbf{k}'_{\perp})^{2}\ell_{B}^{2}\right)\\
 & =\frac{\left(\ell_{B}^{-2}\right)^{2}}{2\pi}\left(-1\right)^{n+m}\frac{m!}{\left(n-1\right)!}\left(\frac{\mathbf{q}_{\perp}^{2}\ell_{B}^{2}}{2}\right)^{n-m}L_{m-1}^{n-m}\left(\frac{\mathbf{q}_{\perp}^{2}\ell_{B}^{2}}{2}\right)L_{m}^{n-m}\left(\frac{\mathbf{q}_{\perp}^{2}\ell_{B}^{2}}{2}\right)\nonumber.
\end{align}
These formulae, of which the validity we have confirmed numerically, are at first sight highly nontrivial.
We guessed them by anticipating that Eqs.~(\ref{eq:PIA1}) and (\ref{eq:pill}) of the main text ought to be equivalent. 
Indeed, this equivalence is realized if and only if Eqs.~(\ref{eq:formula1}) and (\ref{eq:formula2}) are true. 

Armed with Eqs~(\ref{eq:formula1}) and (\ref{eq:formula2}), we arrive at 
\begin{align}
\Pi_V(q) & =-e^{-\frac{\mathbf{q}_{\perp}^{2}\ell_{B}^{2}}{2}}\frac{\ell_{B}^{-2}}{2\pi^{2}}\sum_{nm}\left[C_{nm}^{(1)}(\mathbf{q}_{\perp}^{2})I_{nm}\left(q_{z},i\omega\right)+C_{nm}^{(2)}(\mathbf{q}_{\perp}^{2})J_{nm}\left(q_{z},i\omega\right)\right]\label{eq:PiVVGEN}\\
\Pi_A(q) & =-e^{-\frac{\mathbf{q}_{\perp}^{2}\ell_{B}^{2}}{2}}\frac{\ell_{B}^{-2}}{2\pi^{2}\hbar v_F}i\omega
\Biggl\{\frac{v_Fq_{z}}{v_F^2q_{z}^{2}-\left(i\omega\right)^{2}}\label{eq:PIVAGEN}\\
& +\hbar^2 v_F\sum_{n>0}\left(\frac{\mathbf{q}_{\perp}^{2}\ell_{B}^{2}}{2}\right)^{n}\frac{1}{n!}\sum_{\lambda\lambda^{'}}\int_{-\infty}^{\infty}\text{d}k_{z}\frac{f\left(\lambda \hbar v_F k_{z}\right)-f\left(\lambda'E_{nk'_{z}}\right)}{\left(\lambda\hbar v_Fk_{z}-\lambda'E_{nk'_{z}}\right)^{2}-\left(i \hbar \omega\right)^{2}}\frac{\lambda+\lambda'\frac{\hbar v_Fk_{z}'}{E_{nk'_{z}}}}{2}\Biggr\}\nonumber, 
\end{align}
where
\begin{align}
I_{nm}\left(q_{z},i\omega\right) & =\sum_{\lambda\lambda'}\int\text{d}k_{z}\frac{f(\lambda E_{nk_{z}})-f(\lambda'E_{mk'_{z}})}{\lambda E_{nk_{z}}-\lambda'E_{mk'_{z}}+i\omega}\frac{1+\lambda\lambda'\frac{\hbar^{2}v_F^{2}k_{z}k'_{z}}{E_{nk_{z}}E_{mk'_{z}}}}{2}\\
J_{nm}\left(q_{z},i\omega\right) & =\sum_{\lambda\lambda'}\int\text{d}k_{z}\frac{f(\lambda E_{nk_{z}})-f(\lambda'E_{mk'_{z}})}{\lambda E_{nk_{z}}-\lambda'E_{mk'_{z}}+i\omega}\frac{\lambda\lambda'}{2}\frac{2\hbar^{2}v_F^{2}\ell_{B}^{-2}}{E_{nk_{z}}E_{mk'_{z}}}\\
C_{nm}^{(1)}(\mathbf{q}_{\perp}^{2}) & =\left(\frac{\mathbf{q}_{\perp}^{2}\ell_{B}^{2}}{2}\right)^{n-m}\frac{m!}{n!}\frac{L_{m}^{n-m}\left(\frac{\mathbf{q}_{\perp}^{2}\ell_{B}^{2}}{2}\right)^{2}+\frac{n}{m}L_{m-1}^{n-m}\left(\frac{\mathbf{q}_{\perp}^{2}\ell_{B}^{2}}{2}\right)^{2}}{2}\\
C_{nm}^{(2)}(\mathbf{q}_{\perp}^{2}) & =\left(\frac{\mathbf{q}_{\perp}^{2}\ell_{B}^{2}}{2}\right)^{n-m}\frac{m!}{\left(n-1\right)!}L_{m}^{n-m}\left(\frac{\mathbf{q}_{\perp}^{2}\ell_{B}^{2}}{2}\right)L_{m-1}^{n-m}\left(\frac{\mathbf{q}_{\perp}^{2}\ell_{B}^{2}}{2}\right).
\end{align}
In the derivation of Eq.~(\ref{eq:PIVAGEN}), we have used the fact the transitions from the LL $n$ to $m$ cancel out with
the transitions from $m$ to $n$, if $n$ and $m$ are both nonzero.
When ${\bf q}_\perp=0$, Eqs.~(\ref{eq:PiVVgen}) and (\ref{eq:PIVAgen}) simplify to 
\begin{align}
\Pi_V(q) & =-\frac{\ell_{B}^{-2}}{2\pi^{2}}\left\{ \frac{v_F^2q_{z}^{2}}{v_F^2q_{z}^{2}-\left(i\omega\right)^{2}}+\hbar v_F\sum_{n>0}\sum_{\lambda\lambda'}\int\text{d}k_{z}\frac{f\left(\lambda E_{nk_{z}}\right)-f\left(\lambda'E_{nk'_{z}}\right)}{\lambda E_{nk_{z}}-\lambda'E_{nk'_{z}}+i\omega}\frac{1}{2}\left(1+\lambda\lambda'\hbar^2v_F^2\frac{k_{z}k'_{z}+2n\ell_{B}^{-2}}{E_{nk_{z}}E_{nk'_{z}}}\right)\right\} \label{eq:PiVqpa}\\
\Pi_A(q) & =-\frac{\ell_{B}^{-2}}{2\pi^{2}\hbar v_F}\frac{v_Fq_{z}i\omega}{v_F^2q_{z}^{2}-\left(i\omega\right)^{2}}\label{eq:PiAqpa}.
\end{align}

Equations (\ref{eq:PiVVGEN}) and (\ref{eq:PIVAGEN}) are the main results of this appendix. 
They enable an efficient numerical evaluation of the phonon spectrum in a magnetic field, for the case $b_0=0$. 
The retarded response functions $\Pi^R_V(q)$ and $\Pi^R_A(q)$ are obtained by analytic continuation ($i\omega\rightarrow\omega+i\eta$) of $\Pi_V(q)$ and $\Pi_A(q)$. 

It is useful to make two remarks concerning the numerical evaluation of Eqs. (\ref{eq:PiVVGEN}) and (\ref{eq:PIVAGEN}).
First, we notice that $C_{nm}^{(1)}(\mathbf{q}_{\perp}^{2})=C_{mn}^{(1)}(\mathbf{q}_{\perp}^{2})$ and $C_{nm}^{(2)}(\mathbf{q}_{\perp}^{2})=C_{mn}^{(2)}(\mathbf{q}_{\perp}^{2})$.
Also, for $n\neq0$ and $m\neq0$, we have
\begin{align}
I_{nm}\left(q_{z},i\omega\right)=I_{mn}\left(-q_{z},-i\omega\right) \\
J_{nm}\left(q_{z},i\omega\right)=J_{mn}\left(-q_{z},-i\omega\right) \\
I_{nm}\left(q_{z},i\omega\right)=I_{nm}\left(-q_{z},i\omega\right) \\
J_{nm}\left(q_{z},i\omega\right)=J_{nm}\left(-q_{z},i\omega\right),
\end{align}
which can be proven by shifting  the internal momentum $k_z\rightarrow k_z-q_z$ (for the first two equalities) and $k_z\rightarrow -k_z$ (for the two remaining equalities). 
These relations allow us to gather terms with opposite frequencies ($I_{nm}\left(q_{z},i\omega\right)+I_{nm}\left(q_{z},-i\omega\right)$ for instance) in the sum of Eq.~(\ref{eq:PiVVGEN}). 
This in turn can be exploited to have denominators involving squares of eigenenergies instead of simply energies in Eq.~(\ref{eq:PiVVGEN}). 
As a consequence, the numerical integration over $k_z$ converges better.

Second, a cutoff has to be chosen for the sum over Landau levels. 
This cutoff, denoted $N_\Lambda$, depends on the magnetic field as $N_\Lambda\sim\Lambda^2/(\hbar v \ell_B^{-1})^2$, where $\Lambda$ is a constant. 
The numerical results for the response functions are rather sensitive to the value of $\Lambda$ at low magnetic fields.
At high fields, when Landau levels are well separated and the Fermi energy crosses only a few Landau levels, one can choose $N_\Lambda$ as the highest Landau level intersecting the Fermi energy. 
In this case, adding extra Landau levels to the sum does not change the results qualitatively.

\subsection{Response functions for $b_0\neq0$}

Up to now, we have evaluated $\Pi_A$ and $\Pi_V$ at $b_0=0$. 
It is not difficult to generalize the results from the preceding subsection to the more general case with $b_0\neq 0$. 
We begin by recognizing that, as long as the two Weyl nodes are decoupled, the Green's function $\mathcal{G}_{0}$ is block diagonal in $\tau$. 
Moreover, in each block, $b_0$ can be absorbed by a $\tau$-dependent shift of the chemical potential.
We can thus define node-resolved Green's functions $\mathcal{G}_{0}^\tau$ for each node of chirality $\tau$, by replacing the matrix $\tau_z$ by the number $\tau$ in Eq.~(\ref{eq:SchwinGF}). 
The corresponding node-resolved response function reads
%The response functions $\Pi_V$ and $\Pi_A$ from Eq.~(\ref{PiVVapp1}) can then be expressed respectively as the sum and difference of the single node polarizations
\begin{equation}
\label{eq:pib0}
\Pi_\tau(q)  \equiv \frac{T}{\mathcal{V}}\sum_k\text{tr}\left[\mathcal{G}^\tau_0(k)\mathcal{G}^\tau_0(k+q)\right]= \frac{1}{2}\left[\Pi_V^{(0)}(\mu-\tau b_0)+\tau \Pi_A^{(0)}(\mu-\tau b_0) \right]
\end{equation}
where $\Pi_{VV}^{(0)}(\mu-\tau b_{0})$ and $\Pi_{VA}^{(0)}(\mu-\tau b_{0})$ are the $b_0=0$ response functions (given by Eqs.~(\ref{eq:PiVVGEN}) and (\ref{eq:PIVAGEN})) with chemical potentials equal to $\mu-\tau b_{0}$. 
In the right hand side of Eq.~(\ref{eq:pib0}), we have used the fact that $\Pi_V$ and $\Pi_A$ can be written in terms of the sum and difference of node-resolved response functions, respectively.
%The chiral shift in energy $\tau b_0$ acts only as a shift of the chemical potential with opposite signs for each node. 
As a consequence, $\Pi_V$ and $\Pi_A$ become
\begin{align}
\Pi_{V}	(\mu,b_{0})=\frac{\Pi_{V}^{(0)}\left(\mu-b_{0}\right)+\Pi_{V}^{(0)}\left(\mu+b_{0}\right)}{2}+\frac{\Pi_{A}^{(0)}\left(\mu-b_{0}\right)-\Pi_{A}^{(0)}\left(\mu+b_{0}\right)}{2}\\
\Pi_{A}	(\mu,b_{0})=\frac{\Pi_{A}^{(0)}\left(\mu-b_{0}\right)+\Pi_{A}^{(0)}\left(\mu+b_{0}\right)}{2}+\frac{\Pi_{V}^{(0)}\left(\mu-b_{0}\right)-\Pi_{V}^{(0)}\left(\mu+b_{0}\right)}{2}. 
\end{align} 
These expressions complete the theory of scalar and pseudoscalar response functions in a magnetic field. 
The reader can notice that, since the contribution from chiral Landau levels does not depend on the chemical potential, the results in the quantum limit are not modified by a finite $b_{0}$, unless $|\mu\pm b_0|$ becomes larger than the magnetic gap $\sqrt{n} \hbar v \ell_B^{-1}$.

\section{Comparison with earlier paper [P. Rinkel, P. Lopes and I. Garate, PRL {\bf 119}, 107401 (2017)]}

The aim of this appendix is to compare the results and formalism from this paper to those of our earlier work.\cite{Us} 
In the Supplemental Material of Ref. [\onlinecite{Us}], we obtained the following equation of motion for the phonons:
\begin{align}
\label{eq:former_eom}
\omega^{2} =\omega_{{\rm ph},\mathbf{q}}^{2}-2\frac{\sqrt{N}}{M}{\rm Re}\left[\left(\frac{i\delta\mathbf{Q_{\mathbf{q}}\cdot\mathbf{q}}}{e}\right)g_{0}^{*}\left(\mathbf{q}\right)\right]+\frac{N}{M\mathcal{V}}\frac{V_{\mathbf{q}}}{\varepsilon\left(q\right)}\frac{\left(\mathbf{q}\cdot\mathbf{Q_{\mathbf{q}}^{*}}\right)\left(\mathbf{q}\cdot\mathbf{Q_{\mathbf{q}}}\right)}{e^{2}}+\frac{\mathcal{V}}{M}\Pi_{V}\left(\left|g_{0}\left(\mathbf{q}\right)\right|^{2}+\left|g_{z}\left(\mathbf{q}\right)\right|^{2}\right),
\end{align}
where $\mathbf{Q}_{\mathbf{q}}$ is the total effective phonon charge and
\begin{equation}
\delta\mathbf{Q}_{\mathbf{q}}  \equiv i\frac{e^{2}\mathcal{V}}{\pi^{2}\hbar^{2}\sqrt{N}}\frac{\omega\mathbf{B}}{\omega^{2}-v_F^{2}\mathbf{q}^{2}}g_{z}\left(\mathbf{q}\right)
\end{equation}
is the contribution coming from pseudoscalar phonons in an external magnetic field ${\bf B}$.
In Ref. [\onlinecite{Us}], we knowingly omitted the last term in Eq.~(\ref{eq:former_eom}) by arguing that it did not have a significant impact in the phonon dispersion.
However, we unknowingly missed taking the real part in the second term of the right hand side of Eq.~(\ref{eq:former_eom}) and multiplying it by two; we have properly restored this term here.
Finally, we point out the following change in notation: what we called $\Pi(q)$ in our earlier work is equivalent to $-\hbar v_F\Pi_V(q)/{\bf q}^2$ in the present work.

In Ref. [\onlinecite{Us}], we wrote the total phonon charge as ${\bf Q}_{\bf q}=\mathbf{Q}_{\mathbf{q}}^{(0)}+\delta\mathbf{Q}(q)+\mathbf{Q}_{\mathbf{q}}^{(1)}(q)$, where
\begin{equation}
\mathbf{Q}_{\mathbf{q}}^{(1)}  \equiv i\frac{e\mathcal{V}}{\sqrt{N}}\frac{\Pi_{V}\left(q\right)}{\mathbf{q}^{2}}\mathbf{q}g_{0}\left(\mathbf{q}\right).
\end{equation}
These were the expressions obtained to first order in the external magnetic field. 
In the present work (cf. Appendix A),  we have shown that 
\begin{align}
\mathbf{q}\cdot\delta\mathbf{Q}_{\mathbf{q}} & \equiv i\frac{e\mathcal{V}}{\sqrt{N}}\Pi_{A}\left(q\right)g_{z}\left(\mathbf{q}\right)\nonumber\\
\mathbf{q}\cdot\mathbf{Q}_{\mathbf{q}}^{(1)} & \equiv i\frac{e\mathcal{V}}{\sqrt{N}}\Pi_{V}\left(q\right)g_{0}\left(\mathbf{q}\right)\label{eq:qQ_Pi}.
\end{align}
\begin{comment}
$\Pi_{A}\left(q\right)\equiv\frac{\ell_{B}^{-2}}{2\pi^{2}\hbar v}\frac{\omega v\left|\mathbf{q}\right|\left(\hat{\mathbf{q}}\cdot\hat{\mathbf{B}}_{0}\right)}{\omega^{2}-v^{2}\mathbf{q}^{2}}$
\end{comment}
By inserting Eq.~(\ref{eq:qQ_Pi}) in Eq.~(\ref{eq:former_eom}), we obtain an equation identical to Eq.~(\ref{eq:dispborn}).
This proves the formal equivalence of the phonon equation of motion obtained in different ways in Ref. [\onlinecite{Us}] and the present work.

Despite the formal equivalence, there are important practical differences between our earlier work and the present work. 
In our earlier work, we treated the magnetic field to first order in perturbation theory.
Accordingly, the dielectric function was evaluated at zero field. 
Furthermore we neglected $b_0$ and the chemical potential.
The consequence of these approximations was that the electronic mode hybridizing with the optical phonon came from the pole of the VVA triangle diagram.
In the present work, we have generalized our earlier results to arbitrary strengths of magnetic fields by going to the Landau level picture.
It is interesting that the expression for $\delta{\bf Q}$ from our earlier work agrees almost exactly with the pseudoscalar contribution to the effective phonon charge coming from the chiral LLs. 
The difference lies in the fact that $\Pi_A$ contains a denominator $\omega^2-v_F^2 q_z^2$, instead of the denominator $\omega^2-v_F^2 {\bf q}^2$ found in our earlier work.
Using the expression given in Ref.~[\onlinecite{miransky}] for the Weyl fermion Green's function, the Taylor expansion of $\Pi_A$ to first order in magnetic field reproduces exactly the expression we obtained in the earlier work.
Another difference is that in our earlier work we disregarded the influence of $\Pi_V$ on the phonon dynamics. 
This is no longer a good approximation in the strong magnetic field regime, because $\Pi_V$ develops a singularity at $\omega=v_F q_z$, much like $\Pi_A$ does.
A physical consequence of this is that at strong magnetic fields, the electronic mode hybridizing with the optical phonon is not coming from the pole of the triangle diagram, but rather from the zero of the dielectric function.
In sum, at low fields, the phonon hybridizes with a pseudoscalar boson, whereas at high fields it hybridizes with a plasmon.

To prove that our earlier work was valid within the linear response theory, we will show that $\Pi_V(b_0=0)$ is independent of magnetic field to first order in the field. 
We start from Eq.~(\ref{eq:PiVqpa}) (${\bf q} = q_z \hat{\bf z}$) in the zero temperature limit, and arrive at
\begin{align}
\label{eq:PiVVlowB}
\Pi_V & =\frac{\mu^{2}}{2\pi^{2}\left(\hbar v_F\right)^{3}}\left(\frac{v_F^{2}q_{z}^{2}}{\omega^{2}}\delta_{B}\left[1+2\sum_{n=1}^{\lfloor\frac{1}{\delta_{B}}\rfloor}\sqrt{1-n\delta_{B}}\right]+\mathcal{O}\left(\frac{v_F^4 q_{z}^{4}}{\omega^{4}}\right)\right), 
\end{align}
where $\delta_{B}\equiv 2\hbar^{2}v_F^{2}\ell_{B}^{-2}/\mu^{2}$ and we have considered the regime $v_F q_z\ll\omega\ll\mu/\hbar$ for simplicity.
The first term comes from the chiral LLs and is proportional to $\left|B\right|$
whereas the second term comes from the higher LL intraband terms.
In order to expand the result in the magnetic field, we fix $\delta_B=1/N$ (at which $\Pi_V$ has peaks)
with $N\in\mathbb{N}^{*}$.
This yields
\begin{comment}
\begin{align*}
\Pi_V & =\frac{1}{2\pi^{2}\left(\hbar v_F\right)^{3}}\mu^{2}\left(\frac{v_F^{2}q_{z}^{2}}{\omega^{2}}\frac{1}{N}\left[1+\frac{2}{\sqrt{N}}H_{N-1}^{(-\frac{1}{2})}\right]\right)+\mathcal{O}\left(\mu^{2}\frac{v_F^4 q_{z}^{4}}{\omega^{4}},q_{z}^{2}\right)\\
 & =\frac{1}{2\pi^{2}\left(\hbar v\right)^{3}}\mu^{2}\frac{v^{2}q_{z}^{2}}{\omega^{2}}\left[\frac{4}{3}+\frac{2}{N^{3/2}}\zeta\left(-\frac{1}{2}\right)+\frac{1}{12N^{2}}\right]+\mathcal{O}\left(\mu^{2}\frac{q_{z}^{4}}{\omega^{4}},q_{z}^{2},\frac{1}{N^{\frac{5}{2}}}\right)
\end{align*}
\end{comment}
\begin{align}
\Pi_V & =\frac{2\mu^{2}}{3\pi^{2}\left(\hbar v_F\right)^{3}}\left(\frac{v_F^{2}q_{z}^{2}}{\omega^{2}}\left[1+\frac{3\zeta\left(-\frac{1}{2}\right)}{2N^{3/2}}+\frac{1}{16N^{2}}\right]+\mathcal{O}\left(\frac{v_F^4 q_{z}^{4}}{\omega^{4}},\frac{\hbar^2 v_F^2 q_{z}^{2}}{\mu^2},\frac{1}{N^{\frac{5}{2}}}\right)\right), 
\end{align}
where $\zeta(z)$ is Riemann zeta function.
Hence, the non-analytic contribution ($\sim\delta_{B}$) from the chiral Landau level is canceled at low magnetic field and the amplitude of the peaks in $\Pi_V$ goes as $B^{3/2}$. 
%This dependence is reflected in the quantum oscillations of the sound velocity discussed in the main text.

In the limit $\mu\rightarrow0$, we consider interband contribution
to the polarization only, which yields the known result\cite{Zhou,Thakur}
\begin{align*}
\Pi_V(\mu & =0)=-\frac{1}{6\pi^{2}}\frac{q_z^{2}}{\hbar v_F}\left(\frac{5}{3}+\ln\biggl|\frac{\Lambda^{2}}{\omega^{2}-v_F^2 q_z^{2}}\biggr|-i\pi\Theta\left(\omega-v_F q_z\right)\right)+\mathcal{O}\left(B^{2}\right).
\end{align*}
Here, $\Lambda$ is the ultraviolet energy cut-off and $\Theta$ is the Heaviside function.

Contrarily to $\Pi_V$, $\Pi_A$ has a nonzero linear term in the magnetic field. 
This term is singular at
$\omega=v_F\left|\mathbf{q}\right|$, and corresponds to the anomalous VVA diagram. 
In sum, to first order in the magnetic field, it is justified to approximate $\Pi_V$ by its zero field value, and to keep the chiral anomaly contribution for $\Pi_A$.
This is what was done in our earlier work.

\section{Quantum oscillations in the sound velocity}

In this appendix, we provide some details concerning quantum oscillations in the sound velocity (Fig.~\ref{fig:cs_oscil}).
%scales as $B^{3/2}$ with the magnetic field.
We begin by recognizing that the key quantity governing the oscillations is the real part of $1/\varepsilon$ (see Eq.~(\ref{eq:dispAC_piezo})).
We follow by noticing that, in the regime of interest ($\omega\ll v_F |{\bf q}| \ll \Gamma$ and ${\bf q}||{\bf B}$),
\begin{equation}
\label{eq:1eps}
{\rm Re}\left[\frac{1}{\varepsilon}\right] \simeq -\frac{1}{V_{\bf q} \Pi_{V0}({\bf q},0)}.
\end{equation}
This approximation can be anticipated by using the analytical expression for $\Pi_V$  available in the quantum limit, though we have also verified its validity numerically at weaker magnetic fields where quantum oscillations arise.

According to Eq.~(\ref{eq:1eps}), when $\omega\ll v_F |{\bf q}| \ll \Gamma$ and ${\bf q}||{\bf B}$, the renormalization of the sound velocity is largely determined by the {\em static} and {\em non-disordered} polarization function,
\begin{equation}
\label{eq:pistat}
\Pi_{V0}(q_z,0) = \frac{1}{2\pi^2 \hbar v_F l_B^2}\left[1+2\sum_{n\geq 1} \frac{\sqrt{\mu^2-\Delta_n^2}}{\mu+\Delta_n}+2\sum_{n\geq 1}\frac{\Delta_n}{\hbar v_F q_z} \ln\left(\frac{q_z+2 k_n}{|q_z-2 k_n|}\right)+{\cal O}\left(\frac{\hbar^2 v_F^2 q_z^2}{\Delta_n^2}\right)\right],
\end{equation}
where $\Delta_n=\sqrt{2 n} \hbar v_F l_B^{-1}$ and $\hbar v_F k_n = {\rm Re} \left[\sqrt{\mu^2-\Delta_n^2}\right]$.
Equation~(\ref{eq:pistat}) is valid at zero temperature and for $v_F q_z\ll \Delta_n$ ($\forall n\geq 1$).
The first term in the right hand side (rhs) of Eq.~(\ref{eq:pistat}) is linear in $B$.
We find, numerically, that this linear dependence in the magnetic field is cancelled by the second term in the rhs of Eq.~(\ref{eq:pistat}).
Thus, much like for the regime $\omega\gg v_F q_z$ considered in the previous appendix, the $B$-dependent part of $\Pi_{V0}(q_z,0)$ at weak fields is superlinear.

Equation~(\ref{eq:pistat}) is quite distinct from Eq.~(16) of Ref.~[\onlinecite{Blank}], where a theory of quantum oscillations is presented for conventional metals.\cite{Neuringer}
As a consequence, the magnetic field dependence of the oscillation amplitudes in a Weyl semimetal differs from that of conventional metals.
A detailed study of these differences is left for future work.
%In the latter, the first two 
%In addition, the last term in the rhs of Eq. (\ref{eq:pistat}) contains logarithmic divergences, responsible for the peaks in the quantum oscillations.
%Importantly, the coefficient in front of the logarithm scales as $B^{3/2}$ with the magnetic field.
%This differs from conventional metals, where a $B^{1/2}$ scaling would be expected.\cite{Neuringer}
%At finite but low temperature ($k_B T\ll \hbar v_F l_B^{-1}$), the divergences are reduced to finite peaks.
%The amplitude of the peaks still scales as $B^{3/2}$ with the field, at least when $v_F q_z\ll \Delta_n$.
%This fact, together with Eq.~(\ref{eq:1eps}), explains why the amplitude of the sound velocity has a $B^{3/2}$ dependence on the field.
%{\bf Question: the second term in the rhs of Eq.~(\ref{eq:pistat}) also leads to oscillations, not necessarily with a $B^{3/2}$ dependence, but the amplitude is presumably weak compared to the log terms?}
%{\bf Question: we say that the peaks of the oscillations scales as $B^{3/2}$. Is this scaling valid for the dips too?}

\end{widetext}


\begin{thebibliography}{99}
\bibitem{interm} We assume that $B<B_{QL}$ for all nodes. When $b_0\neq 0$, there exists the possibility that the quantum limit will be reached for one Weyl node but not for the other. The results in this regime do not differ greatly from the ones discussed in the text.
\bibitem{sing} This expression can be derived starting from Eq. (15) and (16) of Ref. [\onlinecite{Thakur}].
\bibitem{caveat} This is an approximate statement. In reality, the contributions involving nonchiral LLs levels are nonzero and they partly screen the piezoelectric coupling even at $\cos\theta=0$. However, sufficiently deep in the quantum limit, these contributions should be small (compared to the contribution made by the chiral LLs at $\theta=0$.)
\end{thebibliography}
\end{document}